\documentclass[nofootinbib,twocolumn,showpacs,superscriptaddress,twoside,prx]{revtex4-1}
\usepackage{amsthm}
\usepackage{amsmath}
\usepackage{amssymb}
\usepackage{graphicx}
\usepackage{url}
\usepackage{dsfont}
\usepackage{float}
\usepackage{bm}
\usepackage{ifthen}
\usepackage[usenames,dvipsnames]{color}
\usepackage{mathrsfs}
\usepackage{float}
\usepackage{afterpage}
\usepackage{subfigure}
\usepackage{adjustbox}
\usepackage{times}
\usepackage{wrapfig}
\usepackage{multirow}

\newcommand{\sket}[1]{{\ensuremath{\lvert#1\rangle}}}
\newcommand{\lket}[1]{{\ensuremath{\left\lvert#1\right\rangle}}}
\newcommand{\tr}[1]{{\text{Tr}}}
\newcommand{\ket}[1]{\if@display\lket{#1}\else\sket{#1}\fi}
\newcommand{\sbra}[1]{{\ensuremath{\langle#1\rvert}}}
\newcommand{\lbra}[1]{{\ensuremath{\left\langle#1\right\rvert}}}
\newcommand{\bra}[1]{\if@display\lbra{#1}\else\sbra{#1}\fi}
\newcommand{\sbraket}[2]{{\ensuremath{\langle#1\rvert#2\rangle}}}
\newcommand{\lbraket}[2]{{\ensuremath{\left\langle#1\!\left\rvert\vphantom{#1}#2\right.\!\right\rangle}}}
\newcommand{\braket}[2]{\if@display\lbraket{#1}{#2}\else\sbraket{#1}{#2}\fi}

\newcommand{\sketbra}[2]{{\ensuremath{\lvert #1\rangle\!\langle #2\rvert}}}
\newcommand{\lketbra}[2]{{\ensuremath{\left\lvert #1\right\rangle\!\!\left\langle #2\right\rvert}}}
\newcommand{\ketbra}[2]{\if@display\lketbra{#1}{#2}\else\sketbra{#1}{#2}\fi}

\newcommand{\RWA}{{\rm{RWA}}}
\newcommand{\od}{{\rm{od}}}
\newcommand{\dd}{{\rm{d}}}
\newcommand{\micro}{{\rm{micro}}}
\newcommand{\drive}{{\rm{drive}}}
\newcommand{\qubit}{{\rm{qubit}}}

\def\({\left(}
\def\){\right)}

\newcounter{lastnote}

\begin{document} 
 
\title{    Universal Quantum  Control through Deep Reinforcement Learning} 

\author{Murphy Yuezhen Niu}
\email{yzniu@mit.edu}
\affiliation{Research Laboratory of Electronics, Massachusetts Institute of Technology, Cambridge, Massachusetts 02139, USA}
\affiliation{Department of Physics, Massachusetts Institute of Technology, Cambridge, Massachusetts 02139, USA}

\author{Sergio Boixo}
\email{boixo@google.com}
\affiliation{Google, 340 Main Street, Venice Beach, California 90291, USA}
 
\author{Vadim Smelyanskiy}
\email{smelyan@google.com}
\affiliation{Google, 340 Main Street, Venice Beach, California 90291, USA}

\author{Hartmut Neven}
\email{neven@google.com}
\affiliation{Google, 340 Main Street, Venice Beach, California 90291, USA}
\date{\today}

\begin{abstract}

Emerging reinforcement learning techniques using deep neural networks have shown great promise in control optimization. They harness non-local regularities of noisy control trajectories and facilitate transfer learning  between tasks. 
To leverage these powerful capabilities for  quantum control optimization, we propose a new control framework  to  simultaneously optimize the speed and fidelity of quantum computation against both leakage and stochastic control errors. 
For a broad family of two-qubit unitary gates that are important for quantum simulation of many-electron systems, we improve the  control robustness by adding control noise into training environments for reinforcement learning agents trained with trusted-region-policy-optimization. 
The agent  control solutions  demonstrate a two-order-of-magnitude reduction in average-gate-error over   baseline stochastic-gradient-descent solutions  and  up to a one-order-of-magnitude reduction in gate time from optimal gate synthesis counterparts.

\end{abstract}

%
%
%
%
%
%
\maketitle
\section*{Introduction}

Designed to exert the full computational force  of Nature, quantum computers      utilize the laws of quantum mechanics  to explore  the exponential computational space in  superposition.  A critical step that connects theory to experiment is the careful design of  quantum   controls  to  translate   each quantum algorithm   into a set of analog control signals that accurately steer  the  quantum computer   around the   Hilbert space.  
The precise choice of these controls ultimately governs the fidelity and speed  of  each   quantum operation.

The fidelity and runtime of  quantum gates are  crucial measures of quantum control that determines the computational capacity of both near- and long-term quantum devices. Higher   gate fidelities lower  the resource overhead for fault-tolerant error correction, while  shorter runtimes  directly extend   quantum circuit depth by racing to avoid the onset of  uncorrectable errors caused by noise and dissipation~\cite{barends2014}.  


Another key component that determines the practical applications of near-term quantum devices is the universality of the quantum gates realizable through analog controls. For pre-fault-tolerant quantum computers,   quantum operations are not limited to   a finite gate  set otherwise necessary for achieving fault-tolerance. Consequently,  implementing high-fidelity and fast  quantum gate  with one control pulse sequence instead of  a long depth circuit through optimal gate synthesis approach can greatly  reduce the resource overhead and expand  the feasible computational tasks. As recently demonstrated in Refs.~\cite{boixo2016characterizing,neill2017blueprint}, replacing the standard  universal gate set  with unrestricted unitary gates   reduces the required  circuit depth for the near-term demonstration of quantum supremacy in experiment by one order of magnitude.


However, a universal control framework that facilitates optimization over major experimental non-idealities under systematic constraints has been lacking, which prevents us from fully leveraging the flexibility of   quantum control schemes.  On the one hand, quantum computing systems with an ever-growing number of qubits are   facing  aggravating amounts of stochastic control errors and information leakage outside the computational subspace. On the other hand, the specific form of system Hamiltonians is limited by the underlying physics of the computing platform and thus unable to   directly induce any  desired quantum dynamical evolution on demand.  Overcoming these challenges  is key to reaping   various computational speedups promised by quantum computers~\cite{feynman1982simulating,grover1996fast,boixo2016characterizing}.

Stochastic control errors  can severely perturb  the  actual control outcomes if not well accounted for during   control optimizations. But in most cases, the exact model of experimental control errors is unavailable.   Efforts towards  improving  control robustness against control errors  have been centered  around closed-loop feedback  optimizations~\cite{dong2010quantum,ruschhaupt2012optimally,chen2013closed,lewis2013reinforcement,palittapongarnpim2017}, which necessitates  frequent measurements of the quantum system.   Since existing experimental   measurements are relatively slow  and can degrade subsequent gate  fidelities,  such closed-loop optimization has yet to become practical for near-term devices.  The majority of  open-loop control optimizations~\cite{nagy2004open,Herschel2014,Wilhelm2015goat}  address   robustness  through the analysis of the  control noise spectrum and control curvature given by the  control Hessian, which quickly becomes computationally exorbitant for  multi-qubit-system  control optimization as system size increases.



Undesirable couplings between a quantum computing system  and its environment also become inevitable when the system is sufficiently large, which induces  information leakage. Such leakage errors  prevent  the implementation of fast   and high-fidelity quantum gates    in many platforms such as   superconducting qubits.  Broadly speaking, two kinds of leakage abound: coherent leakage, which is deterministic and reversible in nature, caused by direct couplings between the qubit subspace and  higher energy subspaces; and `incoherent leakage',   caused by either non-adiabatic transitions\footnote{The non-adiabatic transition  away from the qubit subspace results   from the coherent quantum evolution of the full system, but in the context of the current study such transition is effectively incoherent since  the transition  back  does  not have time to occur.} during  the    modulation of system Hamiltonians  or  by photon loss to the environment.   Coherent leakage  can be further divided   into on-resonant and off-resonant  components,  depending on whether the  frequency components of the control are   close to  the energy gap  separating the  qubit subspace from  a  higher energy subspace~(on-resonant) or not~(off-resonant).

The structure of  high-dimensional  control landscapes of multi-qubit-system  quantum  control problems in the presence of leakage and control errors  are poorly understood due to the lack of analytic tools     and the  prohibitive computational cost  of  numerical approaches. Despite this  lack of precise  knowledge of the control landscape, un-supervised machine learning techniques are able to obtain  high-quality and scalable solutions to similar high-dimensional continuous-variable optimization in real-world problems.  Notably,  reinforcement learning~(RL)    stands out for its usefulness in the absence of labeled data because of its  stability against sample noise and its effectiveness in the face of uncertainty and the stochastic nature of underlying physical systems.  In RL, a  software agent  takes sequential actions aiming to maximize a  reward function, or a negative cost function, that embodies the target problem.  Successful training  of an RL agent depends on balancing  exploration  of unknown territory with  exploitation  of existing knowledge.

Deep RL techniques\cite{schulman2015high,mnih2016asynchronous,silver2016mastering} 
have revolutionized un-supervised machine learning through novel  algorithm designs  which provide scalable,  data efficient, and robust  performance with an  improvement guarantee. Further empowered by advanced optimization techniques using  deep neural networks,   they are able to  solve  more difficult high-dimensional   optimization  problems beyond the reach of classic RL techniques in  benchmark tasks such as  simulated robotic locomotion and Atari games~\cite{schulman2015high,mnih2016asynchronous,silver2016mastering}. While a classic   RL technique, Q-learning,  has  been applied to    quantum control problems recently~\cite{chen2014fidelity,bukov2017machine}, these studies have not yet included practical leakage or control errors. 
We discover in this work that   deep RL techniques are capable of solving more complex   quantum control problems than   previously attempted. The key to leveraging these advanced RL methods is to find an analytic cost function that incorporates the complete objective of the quantum optimization problem.

%
%

%

A comprehensive and efficiently computable  leakage bound of the given  control scheme is one missing piece of a universal control cost function to permit  control optimization for any target unitary gate.   Such lack of an explicit leakage bound also limits the generality of existing  studies.  
For example, Refs.~\cite{khaneja2005optimal,sporl2007optimal,Rabitz2007,moore2008relationship} study quantum controls over all independent  single-qubit Hamiltonians to achieve a provably minimal gate time, but only for closed systems without  leakage.  To minimize on-resonant leakage errors, Ref.~\cite{gambetta2011analytic} turns off independent controls over the  single-qubit Pauli Z couplings,  and Refs.~\cite{martinis2014fast,barends2014,Barry2015high}  turn off single-qubit Pauli X and Y couplings. These hard constraints, however,  could impair  the universality, or the controllability,  over the quantum system: a  time-dependent  evolution without controls over \textit{all} independent single-qubit Hamiltonians is  no longer sufficient to implement an arbitrary unitary gate~\cite{khaneja2005optimal,sporl2007optimal,Rabitz2007}.


%

We propose a control framework, called Universal cost Function control Optimization~(UFO),  to overcome these fundamental challenges in quantum control by connecting deeper physical knowledge of the underlying quantum dynamics with   state-of-the-art RL techniques. 
Instead of resorting from experimental randomized benchmarking for leakage quantification~\cite{wood2017quantification}, we derive an analytic leakage bound for  a Hamiltonian control trajectory  to  account  for both  on- and off-resonant  leakage errors.    
 Our leakage bound  is based on a perturbation theory within the  time-dependent Schrieffer-Wolff transformation~(TSWT) formalism~\cite{Goldin2000} and on a  generalized adiabatic theorem, see  App.~\ref{TSW}.   The use of TSWT is a higher-order generalization of the derivative canceling method for adiabatic gates~\cite{Motzoi2009}, where unwanted leakage errors are suppressed to any desired    order  by adding control Hamiltonians proportional to associated orders of time-derivatives of the dominant  system Hamiltonian. We relax   hard constraints  in control optimization  to soft ones  in the form of adjustable penalty terms of the  cost function, offering more flexibility to an RL agent's   control policy while minimizing the meaningful errors from practical non-idealities.  Our universal cost function enables a joint  optimization over  the accumulated leakage errors, violations of control boundary conditions, total gate time, and gate  fidelity.  Such a framework facilitates   time-dependent controls over \textit{all} independent single-qubit Hamiltonians  and two-qubit  Hamiltonians, thus  achieving    full   controllability~\cite{khaneja2005optimal,sporl2007optimal,Rabitz2007}.

We use the UFO cost function as  a reward for a continuous-variable policy-gradient RL  agent, which is trained by trusted-region policy optimization~\cite{schulman2015high}, to find highest-reward/minimum-cost analog controls for a variety of two-qubit unitary gates. We find that applying  second order gradient methods  to a policy is superior to simpler approaches like direct gradient descent or differential evolution of the control scheme.  We suspect this lies in its ability to leverage non-local features of  control trajectories, which becomes crucial  when the control landscape is high-dimensional and packed with a combinatorially large number of   imperfect saddle points or local optima with vanishing gradients~\cite{dauphin2014}, which is often the case for open quantum systems~\cite{bukov2017machine}.  Moreover, the calculation of  control Hessians is replaced with  a model-free second-order method with  neural networks to further speed  up the optimization process.  In comparison, direct gradient descent methods are known to be incapable of   rapidly escaping such high-dimensional saddle points~\cite{dauphin2014}. 

Our RL agent comprises two  neural networks~(NN): one maps a given  state containing the information about the simulated unitary gate at the current step  to the probability distribution of  proposed  control actions for the next step~(the policy NN); the other maps the same     state  to the projection of the discounted total future reward~(the value function NN)~\cite{schulman2015high}. Both NNs are  fully connected with three layers of dimension 64, 32 and 32 respectively.   Intuitively, the policy NN   encodes the    non-local regularities of numerous  effective control solutions. Such regularities,   traditionally   captured by a carefully chosen analytic functional basis~\cite{Wilhelm2015goat},  are now represented by a model-independent NN without  any   prior knowledge of the mathematical structure  of the target cost function.  The value NN encodes the projected future interactions with  a stochastic environment and  the  associated control cost, which is used to adjust the learning rate of the gradient descent of the policy NN. 

\begin{figure}[h]
\begin{center}
\includegraphics[width=0.8\linewidth]{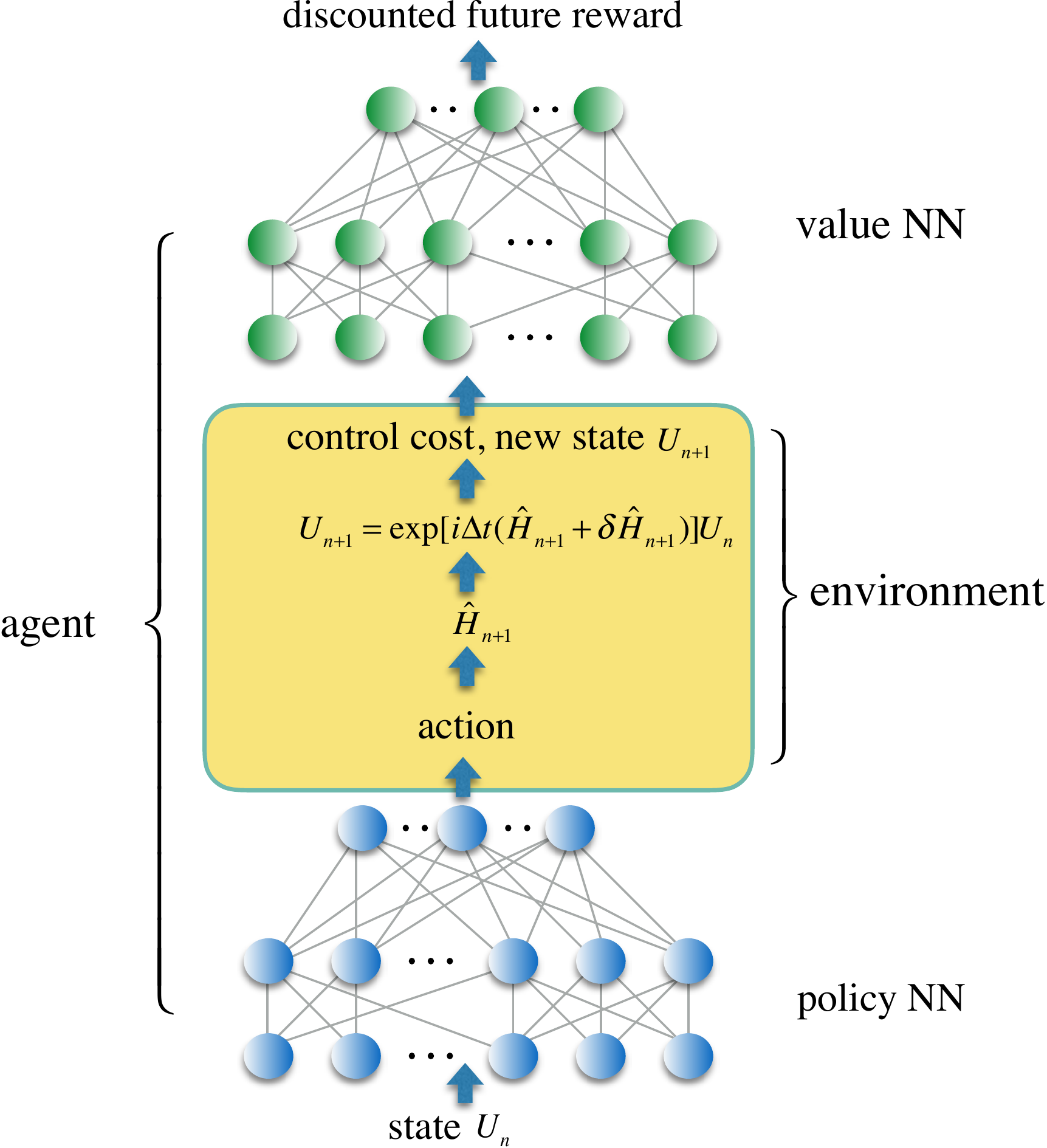}
\caption{Overview of the RL implementation: at the iteration time step $n+1 $,  the   policy NN proposes a control action in the form of the system Hamiltonian $\hat{H}_{n+1}$, the training environment takes the proposed action and evaluates the Schr\"{o}dinger equation under a noisy implementation  $\hat{H}_{n+1} + \delta  \hat{H}_{n+1} $    for time duration  $ \Delta t $ to obtain a new  unitary gate  $U_{n+1}$  and calculates the associated   cost function, both of  which are fed into an RL agent.  The policy NN and value NN of the RL agent are updated jointly based on the trajectory of the simulated unitary gate, control action and associated control cost~\cite{schulman2015trust}.
\label{RLscheme}}
\end{center}
\end{figure}

Both NNs of the RL agent interact  with a training environment that  evaluates the   quantum dynamics under a given control action proposed by the RL agent and returns the updated unitary gate and the corresponding control cost~(as  reward); see Fig.~\ref{RLscheme}.  Optimization consists of many episodes, each of which contains all the time steps  of  a complete quantum control trajectory.  The length of such a sampled control trajectory   is determined   by the minimum of    a predefined  runtime upper bound  and   the time it takes to meet  a termination condition. In our case, the termination condition is measured by a satisfiable value of the  UFO cost function.  After sampling a batch of size $ 20000 $ many  different episodes,  the policy NN is updated to maximize the expected discounted future reward based on the  proposed policy variation within the trusted  region, and the value NN is updated   to fit  the expected discounted future reward based on the newly added samples. A detailed algorithm is presented in \cite{schulman2015high}. We discover that  the  robustness against control errors is significantly improved by simulating experimentally relevant Gaussian-random fluctuations in  control amplitudes through a stochastic RL training environment.  

We verify the quality and the robustness of our control scheme by evaluating the average fidelity of the noise-optimized control solution  under different control noise model parameters. We compare the performance of our RL optimized control solution with the optimal gate synthesis. The latter   provides the minimum number of required gates from a finite  universal gate set for realizing the same unitary transformation. 
 Our RL  control solutions achieve up to a one-order-of-magnitude of  improvement in gate time  over  the optimal gate synthesis approach based on the best known experimental gate parameters in superconducting qubits;   an  order of magnitude reduction in fidelity variance over solutions from both the  noise-free RL counterpart   and a baseline SGD method,  and around  two orders of magnitude reduction in average infidelity over control solutions from the SGD method.


 In the perturbative regime,   the  gmon   Hamiltonian consists of  one-body   and  nearest-neighbor-two-body terms represented by  bosonic creation and annihilation operators, $ \hat{a}_j^\dagger $ and $ \hat{a}_j$, and bosonic number operators  $ \hat{n}_j $, for  each  $ j $-th bosonic mode. In the rotating-wave approximation~(RWA), with a constant rotation rate chosen as the harmonic  frequency of the Josephson junction resonator (see App.~\ref{app:gmon}), the two-qubit gmon  Hamiltonian takes the form:
\begin{align}\label{BSHamiltonian} 
&\hat{H}_{\RWA}(t)=   \frac{\eta}{2} \sum_{j=1}^2 \hat{n}_j(  \hat{n}_j- 1  ) +g(t) ( \hat{a}_2^\dagger\hat{a}_1+\hat{a}_1^\dagger\hat{a}_2) \\\nonumber
&+ \sum_{j=1}^2\delta_j(t) \hat{n}_j  +\sum_{j=1}^2i f_j(t)\left( \hat{a}_j e^{-   i\phi_j(t) } -  \hat{a}_j^{\dagger } e^{ i\phi_j(t) }  \right)\;,
\end{align}
 where the time-independent parameter $ \eta $ represents the anharmonicity of the Josephson junction, and the seven   time-dependent control parameters are:   amplitude $ f_j(t)$ and phase   $\phi_j(t) $ of the microwave control pulse,   qubit detuning $ \delta_j(t) $ with $ j\in\{1,2\} $,  and tunable capacitive coupling or g-pulse $ g(t) $.
The computational subspace is spanned by the two lowest energy levels  of each   bosonic mode: $ \mathcal{H}_{2 }=\text{Span}\{\ket{0}_j,\ket{ 1}_j\}  $, where $ \ket{n }_j $ represents a Fock state with $ n  $   excitations in   the $ j $-th   mode.

%
%

 \section*{Leakage Error Bound}\label{LeakageBoundSec}
To identify different sources of leakage errors, we decompose Eq.~(\ref{BSHamiltonian}) into three parts: $\hat{H}_{\RWA}(t)=\hat{H}_{0} + \hat{H}_{1}(t)+ \hat{H}_{2}(t) $, where $ \hat{H}_{0}= \eta/2 \sum_{j=1}^2 \hat{n}_j(  \hat{n}_j- 1  )  $ accounts for the large constant energy gaps separating the  qubit subspace from other  higher energy subspaces which also determines the minimum  energy gap separating the qubit subspace from the nearest higher energy subspace denoted $ \Delta $,     the block-diagonal Hamiltonian
\begin{align}
&\hat{H}_{1}(t)=\sum_{j=1}^2 \delta_j(t)\hat{n}_j \\\nonumber
&+i f_1(t)  \left( \ket{0}_1\bra{1}_1e^{-i\phi_1(t)} - \ket{1}_1\bra{0}_1 e^{i\phi_1(t)}\right)\otimes \mathbb{I}_2 \\\nonumber
&+i f_2(t)  \,\mathbb{I}_1\otimes \left( \ket{0}_2\bra{1}_2e^{-i\phi_2(t)} - \ket{1}_2\bra{0}_2 e^{i\phi_2(t)}\right) \\\nonumber
&+ g(t)\left(\ket{1}_1\ket{0}_2\bra{1}_2\bra{0}_1 + \ket{0}_1\ket{1}_2\bra{0}_2\bra{1}_1 + h.c. \right)\\\nonumber
&+ g(t)\left(\ket{2}_1\ket{1}_2\bra{2}_2\bra{1}_1 + \ket{1}_1\ket{2}_2\bra{1}_2\bra{2}_1 + h.c. \right)
\end{align} accounts for the  coupling within  the qubit subspace $ \Omega_0=\text{Span}\{ \ket{00},\ket{10},\ket{01},\ket{11}\} $ and within the first excited energy subspace $ \Omega_1=\text{Span}\{  \ket{20},\ket{21},\ket{12},\ket{02} \} $, and the block-off-diagonal  $ \hat{H}_2(t)= \hat{H}_{\RWA}(t)-\hat{H}_{0} - \hat{H}_{1}(t) $  accounts for the  couplings between different energy subspaces with each other. It is the culprit behind  leakage errors,  but since both $ \hat{H}_1(t) $ and  $ \hat{H}_2(t) $ derive from microwave pulses  and the g-pulse, one cannot turn off $ \hat{H}_2(t) $ without turning off controls over single-qubit Pauli X and Y from $ \hat{H}_1 $ which are otherwise crucial for obtaining the full controllability of the qubit system.

In order to suppress and evaluate coherent leakage errors induced by $ \hat{H}_2 $, we   adopt  a  rotated   basis given by the TSWT framework, under the assumption that inter-subspace and intra-subspace couplings are much smaller than the energy gap separating different subspaces:  $  |f_j(t)|\sim |\delta_j(t)| \sim |g(t)|\sim \epsilon \ll \eta \sim \Delta$, see App.~\ref{TSW}. We use $ \epsilon $ and $ \Delta $ to denote the energy scale of the inter/intra subspace coupling strength and the large energy gap separating different energy subspaces, satisfying $\epsilon/ \Delta \ll 1  $.  The  effective block-off-diagonal Hamiltonian $ \hat{\mathbb{H}}_{\od} $ after the TSWT can thus be suppressed to any given higher order by applying the correct order of TSWT.

There are two independent sources of leakage errors for TSWT based quantum control that dominate in superconducting qubit gate controls: the first is the direct coupling leakage  caused by the non-zero block-off-diagonal Hamiltonian after the second order TSWT, and the second is  the leakage caused by the violation of the  adiabaticity due to the fast modulation of the system Hamiltonian.  We derive in App.~\ref{LeakageBoundSec} the bound for the coherent leakage errors as
\begin{align}\label{totalLeakageBoundMaintext}
&L_{tot}= \frac{ \Vert \hat{\mathbb{H}}_{\od}(0) \Vert }{\Delta(0)}\\\nonumber
&+ \frac{ \Vert \hat{\mathbb{H}}_{\od}(t) \Vert }{\Delta(t)} +  \int_0^t \frac{1}{\Delta^2(t)} \left|\left| \frac{d^2 \hat{\mathbb{H}}_{\od}(t)}{dt^2} \right|\right|dt,
\end{align}
where $ \hat{\mathbb{H}}_{\od} $ represents the block-off-diagonal Hamiltonian, which is of magnitude $ O(\frac{\epsilon^3}{\Delta^2} )$ after the second order TSWT.


In addition to the coherent leakage errors bounded  by   Eq.~(\ref{totalLeakageBoundMaintext}) there also exists incoherent leakage errors due to the violation    of adiabaticity from the time-dependent nature of our control quantum dynamics in the off-resonant regime.   We derive   a generalized adiabatic theorem  to bound  the non-adiabatic leakage error; see App.\ref{LeakageBoundSec}. We  show that such  non-adiabatic leakage  is not dominated in off-resonant frequency regime and   Eq.~(\ref{totalLeakageBoundMaintext}) accounts for dominant leakage errors in both on-resonant and off-resonant regimes.

\section*{Universal  Cost Function}
 An effective control cost function  is crucial  to  an efficient   control optimization   and to guaranteeing   the full controllability over the quantum system. We   propose    a  control cost function that  includes leakage errors, control constraints, total runtime, and gate infidelity as soft penalty terms that are readily optimizable using RL techniques without compromising   the system controllability.   We illustrate the design of a UFO cost function in the  tunable gmon superconducting-qubit  architecture~\cite{Chen2014}. 
 
 A   unitary gate is realizable through the control of the time-dependent Hamiltonian defined  in  Eq.~(\ref{BSHamiltonian}) according to  $ U(T)=\mathbb{T} [\exp(-i\int_0^T\hat{H}_{\RWA}(t) dt)]$, with $ \mathbb{T} $ denoting the   time-ordering operator.  The inaccuracy of the controlled two-qubit  unitary gate $  U(T)$  with respect to a  target unitary gate $ U_{\text{target}} $ is measured by the gate infidelity: $ 1-F[U(T)]= 1-1/16\left| \text{Tr} ( U^\dagger(T)U_{\text{target}})\right|^2 $ \cite{khaneja2005optimal,sporl2007optimal,Rabitz2007,moore2008relationship},  which vanishes  only when  $ U(T)=U_{\text{target}} $ up to a global phase. This definition of control inaccuracy  is widely used in quantum control optimization~\cite{khaneja2005optimal,sporl2007optimal,Rabitz2007,Motzoi2009,bukov2017machine, martinis2014fast,barends2014,Barry2015high} for its  little  accompanying  computational overhead   during iterative optimizations.    Gate infidelity can also be bounded by the average gate infidelity measured through experimental benchmarking: lower gate infidelity implies lower average gate infidelity~\cite{sanders2015bounding}.  For these two reasons, we  choose gate infidelity as  the first part of our UFO cost function to penalize the control inaccuracy.   The second part  is a penalty term on the accumulated  leakage errors derived above.  The last two terms of the control cost function  penalize  the total runtime $ T $ and the violation of  control boundary conditions. Boundary conditions are chosen     to facilitate convenient gate concatenations:  microwave pulses and the   g-pulse should   vanish at both boundaries such that the computational bases and the  Fock  bases coincide. This is  enforced   by adding    $\sum_{t\in \{0,T \}} [  g(t)^2  +  f(t)^2]$ to  the control cost function. Such boundary constraints also help  to minimize the errors caused by   deviations from RWA due to the fast oscillating nature of the non-RWA terms; see  App.~\ref{app:gmon}.  We thus obtain the  full  UFO cost function:
\begin{align}\nonumber
&C(\chi, \beta,\gamma, \kappa)= \chi [1-F[U(T) ] ]+ \beta L_{\text{tot}} \\
& + \mu \sum_{t\in \{0,T \}} \left[   g(t)^2 +  f(t)^2  \right] + \kappa T\label{totalCost}
\end{align}
where $\chi$ penalizes the  gate infidelity,  $ \beta$ penalizes different sources of   leakage errors,   $  \mu$ penalizes  the violation of   boundary constraints,  and $ \kappa$ penalizes the total runtime. These hyper-parameters are optimized to balance the joint optimization for achieving  satisfactory control outcomes.  To apply to other quantum computing platform where our control constraints no longer applies,  each  term of the UFO cost function can be modified to   best describe optimization target based on  the underlying physics.

\section*{Two-Qubit Gate Control Optimization}
 
We now apply the UFO framework to find fast and high-fidelity  two-qubit gate controls  that are  robust  against  control errors.  We define the robustness of a  gate control under a given control noise model   as a bounded  deviation of the average quantum gate fidelity $ \bar{F}( \mathcal{E} ,U_{\text{target}}) $ from  an  ideal average gate   fidelity $ F_{\text{ideal}} $:
\begin{align}
 | \bar{F}( \mathcal{E} , U_{\text{target}})  -F_{\text{ideal}}  |  < \epsilon_0 , \,\, \text{for} \, \, \epsilon_0 >0,
\end{align}
where   
\begin{align}
\bar{F}( \mathcal{E} , U_{\text{target}})= \int d\psi\bra{\psi}U_{\text{target}}^\dagger \mathcal{E}(\ket{\psi}\bra{\psi})U_{\text{target}}\ket{\psi} 
\end{align} embodies the quality of the gate-control quantum channel by  averaging  over the whole state space under a uniform Haar  measure\cite{nielsen2002simple}, with the trace-preserving quantum  operation $ \mathcal{E} $ accounting for the    noisy implementation of a target unitary $  U_{\text{target}} $; (see App.~\ref{aveFidapp}). The average gate infidelity is    defined accordingly as $ 1- \bar{F}( \mathcal{E} , U)$.

Such a robustness criterion can be validated for a given control scheme using a number of computational steps that is linear in  the total degrees of freedom of control parameters. However, it differs  from  the canonical definition in optimal control theory~\cite{nagy2004open,Herschel2014}, where the   number of computational steps for the analysis of   robustness using control Hessians   scales  cubically  with the total degrees of freedom in control parameters. For special cases, such as closed-system single-qubit control,  there exist   analytic expressions for the control Hessian~\cite{nagy2004open,Herschel2014}.  But in the current work  we choose a more practical  definition of robustness scalable  to  multi-qubit control problems. 


Traditional quantum control trajectory optimization  depends on the  complete  knowledge of the underlying physical model. In contrast, the success  and robustness of RL persist with incomplete and  potentially flawed modeling.  It is often the case in experiments that the exact   control error model  is unknown. Given partial information about the control error model, can we leverage RL optimization to  find robust control solutions against  not just one but a   set of control error models?
In our case,  we deploy  RL agents trained by trust-region policy optimization~\cite{schulman2015high} in the OpenAI platform~\cite{brockman2016openai}, to find near-optimal control solutions to the UFO cost function described in Eq.~(\ref{totalCost}).   We incorporate a pertinent  control noise model of gmon superconducting-qubit Hamiltonian~\cite{Chen2014} into  a stochastic training environment.  At each time step,  amplitude fluctuations   sampled from a  zero mean Gaussian distribution with 1 MHz variance, which amounts to around   $ 5\% $   control parameter uncertainty, are added to  Hamiltonian  parameters   that are known to be prone to   fluctuations:  qubit anharmonicity,  qubit detuning amplitudes, microwave control amplitudes and qubit  g-pulse  amplitude, see App.~\ref{app:gmon}.  Harnessing the  sample-noise resilience  of  RL optimization, we expect the optimized control to be robust against a family of control noise models despite being trained under a single model. This  is indeed  proven to be   the case as evidenced  by our numerical simulations.


\begin{align}
\mathcal{N}(\alpha, \alpha,  \gamma)=\exp[i(\alpha \sigma_1^x\sigma_2^x+ \alpha \sigma_1^y\sigma_2^y+\gamma\sigma_1^z\sigma_2^z)
\end{align}

\begin{figure}[h]
\begin{center}
\includegraphics[width=0.8\linewidth]{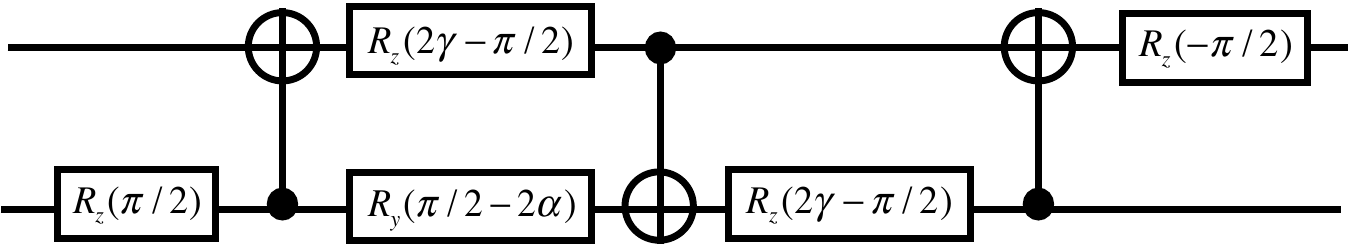}
\caption{Optimal gate synthesis for realizing unitary gate $ \mathcal{N}(\alpha,\alpha, \gamma) $.
\label{optcirc}}
\end{center}
\end{figure}

In    gmon superconducting qubits, the energy gap that separates the qubit subspace from the nearest higher energy subspace is $ \Delta(s)\approx 200 $ MHz.  We apply control  frequency  filters (App.~\ref{app:cfd}) to  piece-wise constant analog control  signals such that the   frequency  bandwidth of the proposed Hamiltonian modulation is limited to be within  $ 10 $ MHz. Given that our  off-diagonal Hamiltonian after the second order TSWT  is of order $ 1/10 $ MHz~(App.~\ref{TSW}),  the first term of the leakage bound  in Eq.~(\ref{totalLeakageBoundMaintext} ), $  \int_0^1 \frac{1}{\Delta^2(s)}\frac{1}{T}\left|\left| \frac{d^2 \hat{\mathbb{H}}_{\od}(s)}{ds^2} \right|\right|ds $, is  of order $ 10^{-4} $, around the fault-tolerant threshold value for leakage error of near-term surface code~\cite{Fowler2015}.   Although the gmon Hamiltonian is fully controllable under our UFO paradigm, we target at a family of two-qubit gates parametrized by  $ \mathcal{N}(\alpha, \alpha,  \gamma)=\exp[i(\alpha \sigma_1^x\sigma_2^x+ \alpha \sigma_1^y\sigma_2^y+\gamma\sigma_1^z\sigma_2^z)]\ $,
where $ \sigma_j^k $ is the $ k\in \{X, Y, Z\} $ Pauli matrix of the $ j $-th qubit. The  optimal gate synthesis~\cite{Colin2004} that provides the   optimal decomposition of such unitary transformation into a  minimum number of arbitrary single-qubit rotations and CZ gates corresponds to a depth seven circuit containing   three two-qubit gates and five single-qubit gates, see Fig.~\ref{optcirc}. This gate family  includes    the  SWAP, ISWAP, CNOT and CZ, fermionic swap gate,  and Given's rotation up to single-qubit rotations. Both the fermionic swap gate  and Given's rotations  are  used  for  realizing Jordan-Wigner transformations in fermionic Hamiltonian simulation~\cite{Troyer2015,kivlichan2017quantum,jiang2017quantum}.  Identifying   continuous controls that outperform their optimal gate synthesis  counterparts for this family of gates thus has  far-reaching   applications across  quantum chemistry and   quantum simulation. The hyper-parameters of the UFO cost function are optimized through a grid search and is applicable to  all target gates we have considered: $ \chi=\beta=10,  \mu=0.2, \kappa=0.1 $.


We compare our noise-optimized control  obtained by the RL agent with the optimal gate synthesis counterpart in overall runtime. Based on state-of-the-art experimental   implementations, we set the gate time for each single-qubit gate to $ 20 $ns and CNOT to $ 45 $ns.  The   optimal gate synthesis in Fig.~\ref{optcirc} thus takes      $ 215 $ns in  runtime.


\begin{figure}[h]
\begin{center}
\includegraphics[width=1\linewidth]{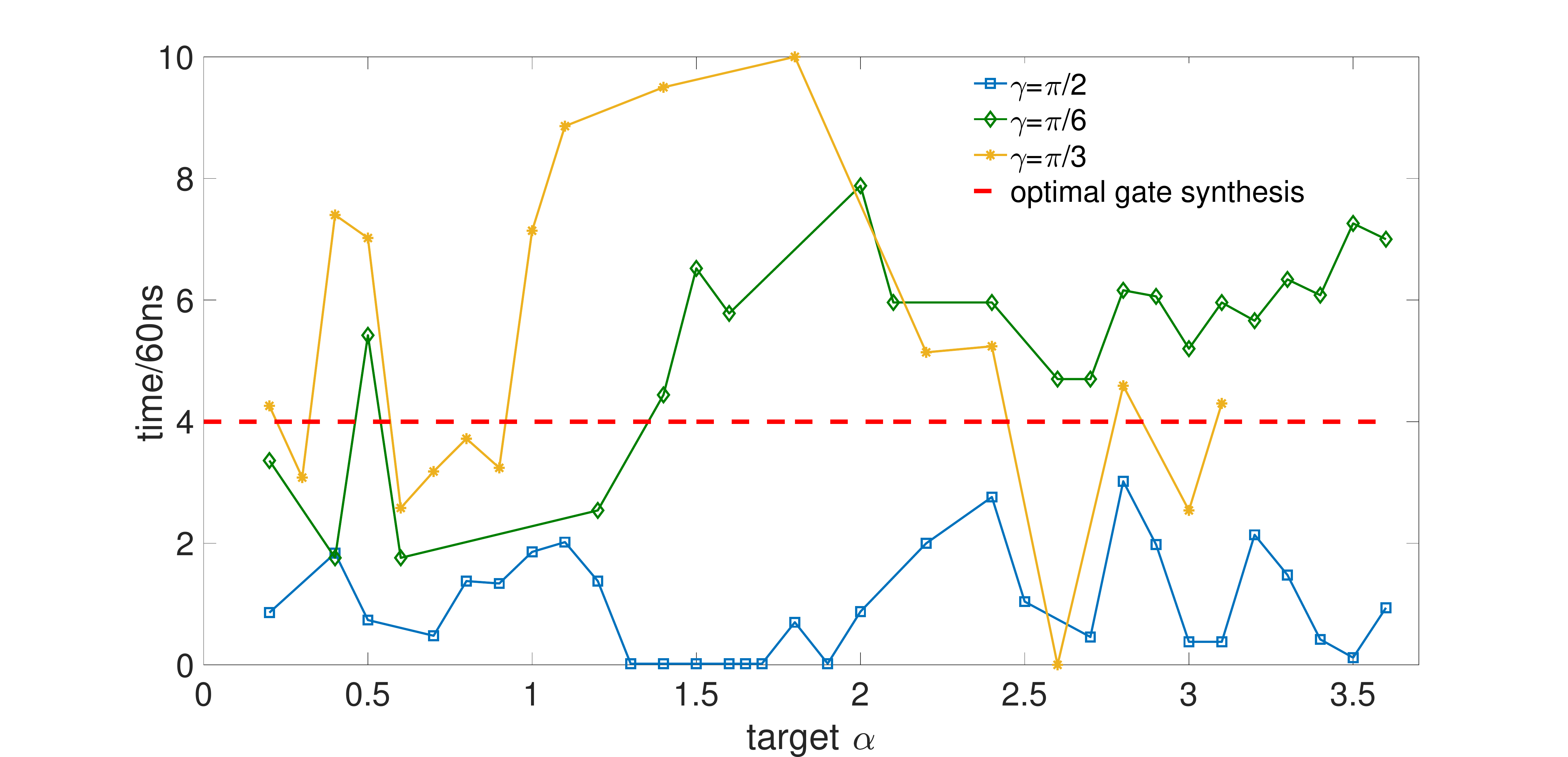}
\caption{Gate run time of two-qubit gate family $ \mathcal{N}(\alpha,\alpha, \gamma) $ for $ \gamma=\pi/2 $~(blue curve),  $ \gamma=\pi/6 $~(green curve) and  $ \gamma=\pi/3 $~(yellow curve). The standard optimal gate synthesis run time for this gate family is around $ 200 $ns marked by dashed red line. Total leakage errors and gate infidelity  are upper bounded by   $O(10^{-4})$ and $O(10^{-3})$ respectively for all cases. 
\label{TimeComp}}
\end{center}
\end{figure}

The  gate times of our noise-optimized control schemes for three different values of $ \gamma $ are  shown in Fig.~\ref{TimeComp}.  There, different data points of the same $ \gamma $ are obtained by the same RL agent with an adaptive step size in $ \alpha $ to guarantee a constant upper bound on the total optimization time: target gate  $ \alpha $ will be increased by one step $ \alpha= \alpha + 0.1 $,  either when  the agent obtains a control solution with a low enough overall cost,   or when the   optimization time for a given  $ \alpha $  exceeds a pre-defined  value. We discover that it takes significantly less time for an RL agent  to learn a new target unitary gate  based on the successful learning of  a nearby target   than   to learn a new target gate  afresh, which provides heuristic evidence for the   transfer learning  facilitated by RL using deep NN\footnote{The use of adaptive step size can be replaced by parallel RL agents, each dedicated to a fixed target unitary gate, which is not the focus of  the  current study.}.

We have seen  a factor of 10 runtime improvement  for the two-qubit gate family parametrized by $  \mathcal{N}(\alpha,\alpha, \pi/2)  $  with  $   \alpha \in [0, \pi] $ over the optimal gate synthesis. Such significant improvement manifests the hardware efficiency of our control optimization: the target unitary gate can be  rewritten as  $  \mathcal{N}(\alpha,\alpha, \pi/2) =-\exp[i(\alpha \sigma_1^x\sigma_2^x+ \alpha \sigma_1^y\sigma_2^y)]\exp[-i\frac{\pi}{2}\sigma_1^z]\exp[-i\frac{\pi}{2}\sigma_2^z] $  whose two-qubit entangling part is  directly  realizable through  a time evolution under the gmon Hamiltonian defined in Eq.~(\ref{BSHamiltonian}) without   detuning or microwave controls: $ \delta_j(t)=f_j(t)=0 $ with $ j\in\{1,2\} $.  Our RL control optimization is thus able to detect such an inherent regularity, which  relates   a given system Hamiltonian to the family of  target unitary gates that are efficiently implementable.  Isolated peaks in the gate time plot in Fig.~\ref{TimeComp} are potentially due to control singularities, which suggests the need for  further studies into the hardness of the   analog-control  landscape in the presence of leakage and control errors.

We verify  the robustness of the noise-optimized  control solution    $ \vec{c} $  from RL by evaluating  its  average fidelity $ \bar{F}(\mathcal{E},U_{\text{target}}) $ and  the  variance of the  control gate  fidelities $ F[U(\vec{c}) ] $   under different control noise  instances   $ \delta \vec{c}  $   sampled from the same Gaussian distribution $N(0,\sigma_{\text{noise}})  $:
\begin{align}  
&\sigma_{\text{fidelity}} = \mathbb{E}_{ \delta \vec{c} \sim  N(0,\sigma_{\text{noise}}) }\left( F[U(\vec{c} +\delta \vec{c}) ]-F_{\text{ave}} \right)^2,\\
& F_{\text{ave}}= \mathbb{E}_{ \delta \vec{c} \sim N(0,\sigma_{\text{noise}}) } F[U(\vec{c} +\delta \vec{c}) ].
\end{align}


\begin{figure*}[ht]
\centering
\includegraphics[scale=0.3]{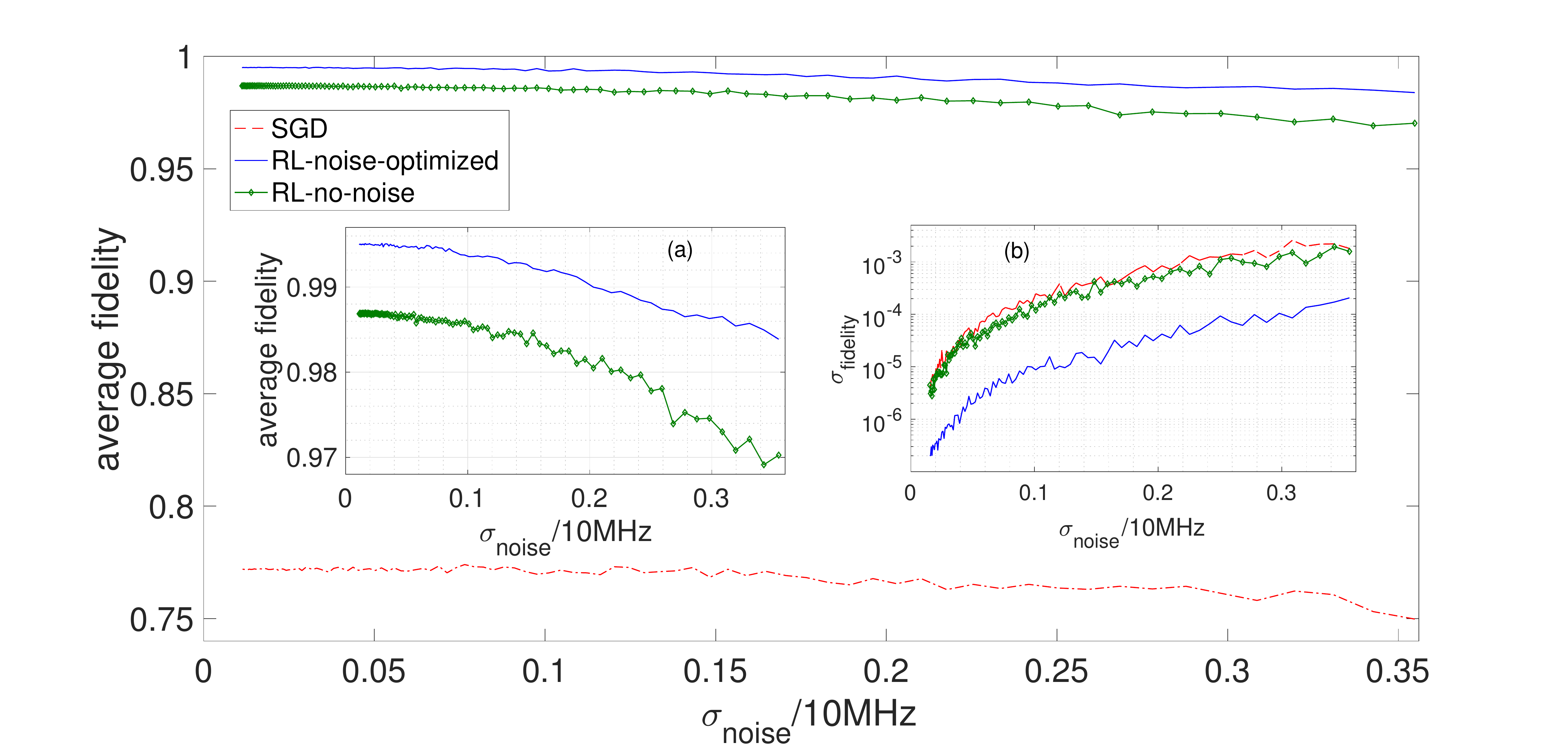}
\caption{Average fidelities of the optimized quantum control schemes vs the Gaussian control noise  variance for the gate $  \mathcal{N}(2.2, 2.2, \pi/2) $.   The blue line  represents the performance  of the noise-optimized  control obtained by an RL agent trained under a noisy environment. The green line marked by diamond shapes represents the performance of the control obtained by an RL agent with a  noise-free environment. The red dashed line  represents the performance of the control trajectory obtained by SGD.  Subplot (a): zoomed in comparison of the average fidelities  of the noise-optimized and noise-free RL control solutions under different values of   Gaussian control   noise  variance.
 Subplot (b):   comparison of fidelity variances of three different control schemes  under different    control noise variances $ \sigma_{\text{noise}} $, where each data point is taken from    60 different control trajectories with control amplitude error at every time step sampled from the Gaussian distribution $ N(0, \sigma_\text{noise}) $.
\label{fidNoise}}
\end{figure*}

We consider a Gaussian family of stochastic control error models:  the  amplitude fluctuations of control parameters  are  described by  Gaussian  distributions of  zero mean  and a variance $ \sigma_{\text{noise}} $  ranging from $ 0.1 $ MHz to $ 3.5 $ MHz.  The gate control performance under the noise model with   $1$ MHz variance is a reasonable indicator for experimental implementations.  Nevertheless, the exact value of control amplitude variance is   hard to determine and can drift over time. 
The blue curve    in Fig.~\ref{fidNoise} represents the average fidelity of the noise-optimized control by RL, which   stays within the  range of $ [99.5\%, 98\%] $ under the given noise model parameter range, satisfying our control robustness definition with $ \epsilon_0=0.007 $ at $ \sigma_{\text{noise}}=1 $MHz. 
 In Fig.~\ref{fidNoise}, we compare noise-optimized control with a noise-free control solution obtained by  an RL agent without a stochastic environment, represented by the green curve marked by diamonds, and with that obtained by a  baseline stochastic gradient descent~(SGD) technique using the Adam optimizer~\cite{kingma2014adam}, represented by the red dashed lines.  The noise-optimized control solution    manifests up to a one-order-of-magnitude  improvement  in average gate infidelity over the noise-free control solution using RL, and around two-order-of-magnitude improvement in   average gate infidelity     over   SGD baseline solutions.  Moreover, the sampled fidelity variance of the noise-optimized RL solver is  one  order of  magnitude lower than two   other  methods consistently throughout the tested noise model parameter range. This validates the improved stability of our control solution obtained by a policy-gradient trained  RL agent against experimentally relevant Gaussian control noise models.

\section*{Conclusion}

We propose a quantum control   framework, UFO, for fast and high-fidelity quantum gate control optimization. It is   applied to an open-loop control optimization through reinforcement learning, where the control trajectory is encoded by a first neural network (NN) and the control cost function is encoded by a second NN. Robust control solutions are  obtained by training both NNs under a stochastic environment mimicking   noisy control actuation.  We achieve up to one order of magnitude reduction  in average gate infidelity over noise-free alternatives and up to a one-order-of-magnitude reduction in gate time over the optimal gate synthesis solution.
This is significant, given that  the  highest gate fidelity in  state-of-the-art  superconducting-qubit systems is around $99.5\%$, and that the  total computation runtime is limited by decoherence to several microseconds.

Our work opens a new direction  of quantum analog control optimization using RL,  where unpredictable control errors and  incomplete physical models of environmental interactions are  taken into account during the control optimization.  Other  advanced machine learning techniques are also readily applicable to our control  framework. The success of deep RL   in Alpha Go~\cite{silver2016mastering}  and   robotic control~\cite{schulman2015high,mnih2016asynchronous} suggests that our approach--once  generalized to closed-loop control optimization, where system calibration and gate control optimization are combined into a unified procedure--could further improve control robustness towards systematic and time-correlated errors. 

\paragraph*{Acknowledgement}

We thank Charles Neil and Pedram Roushan for  helpful discussions on the experimental noise models and detailed control parameters   in the gmon superconducting qubit implementation. M.~Y.~N. thanks   Barry  C. Sanders for early discussions on control optimization through reinforcement learning,    Isaac L. Chuang for useful comments on the draft,   Th\'{e}ophane Weber, Martin McCormick and Lucas Beyer
 for   discussions on  reinforcement learning methods.
 M.~Y.~N. acknowledges financial support from the Google research internship summer 2017  program during which the majority of the work was performed.  M.~Y.~N.  acknowledges support from the Claude E. Shannon Research Assistantship.   

\paragraph*{Author Contributions.}  M.~Y.~N, S.~B. and V.~S. developed the project and theory. M.~Y.~N. implemented the reinforcement algorithms and numerical studies. All authors contributed to the draft.

\onecolumngrid
\appendix

\begin{appendix}

\section{Gmon Hamiltonian and Control Noise }\label{app:gmon}

We focus on the two-qubit gmon Hamiltonian in the perturbative regime as an example. It is  obtained by quantizing a phenomenological classical Hamiltonian for non-linear  L-C circuits containing Josephson junctions, as explained below.

\subsection{Quantum Oscillator Picture}

The supercurrent across a Josephson junction between two superconducting media, in the absence of any external field, persists and undergoes quantum oscillations once connected to a capacitor. Intuitively, it is the quantum tunneling across the junction that drives the supercurrent. Such macroscopic quantum effect is associated with a phenomenological classical circuit model, where the effective voltage across the Josephson junction comes from the changing of the phase difference $ \phi $ across the junction as
\begin{align}
V=\frac{\phi_0}{2\pi}\frac{d \phi}{d t}\;,
\end{align}
where $ \phi_0=h/2e $ and $ e $ is the electron charge. Moreover, the supercurrent depends on the phase difference as
\begin{align}
I=I_c\sin(\phi)\;,
\end{align}
where the critical current $ I_c $ is a parameter characterizing the junction. Once we connect it to a capacitor of capacitance $ C $, we obtain a nonlinear quantum oscillation. We write the potential energy of the inductor as
\begin{align}\label{PotentialJJ}
U&=\int IVdt =\frac{\phi_0 I_c}{2\pi}\int  \sin(\phi) \frac{d \phi}{d t}dt=-E_J \cos\phi\;.
\end{align}
The  energy  contribution from the capacitor
\begin{align}\label{capacitorE}
T=\frac{1}{2}CV^2=\frac{1}{2}C\(\frac{\phi_0}{2\pi}\dot{\phi}\)^2
\end{align}
can be regarded as the kinetic term of the oscillator energy. The Hamiltonian for the Josephson junction oscillator, to lowest order in $ \phi $, is 
\begin{align}
H_0=U_0 + T = \alpha \phi^2 + \beta \dot{\phi}^2\;,
\end{align}
with $ \alpha $  and $ \beta $ determined by the effective capacitance and inductance of the superconducting circuit. For example, we can add additional inductor into the Josephson L-C circuit to change the potential energy of the effective inductance of the whole circuit.

To analyze the quantum dynamics of this phenomenological model derived from its classical analogue, we perform second quantization of $ \phi$  by treating it as one of the two quadratures of a bosonic field:
\begin{align}
 \hat{\phi}=\kappa \frac{ \hat{a}^\dagger+\hat{a} }{\sqrt{2}}\;,
\end{align}
where $ \kappa=(\beta/\alpha)^{1/4} $. 
The conjugate basis $\dot{ \hat{\phi}} $ takes the  form
\begin{align}\label{conjugatePhase}
\dot{ \hat{\phi}} = \kappa^{-1}\frac{ \hat{a}^\dagger-\hat{a} }{\sqrt{2}i}\;.
\end{align}
Under this quantization, the harmonic part of the Josephson L-C circuit Hamiltonian is
\begin{align}
\hat{H}_0=\omega_0 \(\hat{a}^\dagger\hat{a} + \frac{1}{2})\)\;,
\end{align}
where we have set $ \hbar=1 $ and defined $ \omega_0=\sqrt{\kappa} $ . 

Now let us include the lowest order anharmonic term from Eq.~(\ref{PotentialJJ}) to  the Josephson  L-C circuit Hamiltonian:
\begin{align}
\hat{H}_1&= \gamma \hat{\phi}^4=\frac{\gamma \kappa^4}{4}(\hat{a}+\hat{a}^\dagger)^4\\\label{beforeRWA}
&=\frac{\eta}{8}\left[\left(\hat{a}^2+(\hat{a}^\dagger)^2\right)^2 +\left(\hat{a}^2+(\hat{a}^\dagger)^2\right)  (2 \hat{n}+1)  +  (2 \hat{n}+1) \left(\hat{a}^2+(\hat{a}^\dagger)^2\right) + (2 \hat{n}+1)^2  \right]
\end{align}
with $ \hat{n}=\hat{a}^\dagger\hat{a} $, $ \gamma  $ determined by the configurations of additional inductor or capacitor elements added into the Josephson L-C circuit and $ \eta $ as a   parametrization following literature convention.

The coupling between two qubits is realized by  turning on the capacitive coupling between two   Josephson L-C circuits.  To the lowest order, such capacitive coupling  energy depends on  $ \dot{\phi_1} $ and $ \dot{ \phi_2 }$ as:
\begin{align}
\hat{H}_{2}= -\alpha^\prime(t)  \dot{\phi_1} \dot{ \phi_2 }=-\frac{\alpha^\prime(t)}{2\kappa_1\kappa_2}(\hat{a}_1^\dagger-\hat{a}_1)(\hat{a}_2^\dagger-\hat{a}_2)\;,
\end{align}
where the capacitive coupling strength $ \alpha^\prime(t)$ is tunable.

Additionally, we also have capacitive coupling  between  a strong non-depleted microwave pulse   and our quantized bosonic modes  as
\begin{align}
T_{\micro}=\frac{1}{2}CV_{\drive} V_{\qubit}\;.
\end{align} 
The microwave drive induced voltage  is proportional to the field amplitude
\begin{align}\label{micdrive}
V_{\drive}=f(t)e^{-i\omega t - \phi} + f(t)e^{i\omega t+ \phi}
\end{align} 
with  c-number $ F(t) $ determined by the amplitude of the microwave drive, $ \phi $ being the phase of the drive and $ \omega $ being the frequency of the drive. The quantized voltage of the bosonic field, on the other hand,  can be represented by the conjugate  phase operator  defined in Eq.~(\ref{conjugatePhase}) 
\begin{align}
V_{\qubit}= i  ( \hat{a}_j-  \hat{a}_j^\dagger)
\end{align}
up to a real-valued constant factor. Putting together the voltage dependence on the control field and quantized bosonic modes, we obtain the microwave pulse controlled   capacitive energy:
\begin{align}
\hat{H}_3=\sum_{j=1,2}i \left( \hat{a}_j-  \hat{a}_j^{\dagger }  \right)f_j(t)\cos(\omega_j t +\phi_j)\;,
\end{align}
where   $ \omega_j $, $ \phi_j $,  and $ f_j(t) $ represents respectively the frequency, phase and amplitude of each microwave pulse. The overall system Hamiltonian thus comprises four parts
\begin{align}
\hat{H} &=\hat{H}_0 + \hat{H}_1 + \hat{H}_2+ \hat{H}_3.
\end{align}
We further simplify this Hamiltonian by  switching to the interaction picture
\begin{align}
 \hat{H}_I= U\left( \hat{ H}- \hat{\tilde{H}}_0 \right) U^\dagger
\end{align}
with $ U=e^{-i\hat{\tilde{H}}_0 t} $  and
\begin{align}
\hat{\tilde{H}}_0=  (\bar\omega_0 +\eta) \hat{n} =\tilde{ \omega}_0\hat{n}\;.
\end{align}
The  Hamiltonian for two bosonic modes  hosted by two coupled gmon circuits in the interaction picture thus takes the form
\begin{multline}
 \hat{H}_I=\sum_{j=1}^2\delta_j \hat{n}_j+ \frac{\eta}{8}\sum_{j=1}^2[\left(\hat{a}_j^2e^{-2i\tilde{\omega}_0 t}+(\hat{a}_j^\dagger)^2e^{2i\tilde{\omega}_0 t}\right)^2 +\left(\hat{a}_j^2e^{-2i\tilde{\omega}_0 t}+(\hat{a}_j^\dagger)^2e^{-2i\tilde{\omega}_0 t}\right)  (2 \hat{n}_j+1)\\
  +  (2 \hat{n}_j+1) \left(\hat{a}_j^2e^{-2i\tilde{\omega}_0 t}+(\hat{a}_j^\dagger)^2e^{2i\tilde{\omega}_0 t}\right) + 4 \hat{n}_j^2 -4 \hat{n}_j+ 1  ]\\
-\frac{\alpha^\prime(t)}{2\kappa_1\kappa_2}(\hat{a}_1^\dagger\hat{a}_2^\dagger e^{2i\tilde{\omega}_0 t}+\hat{a}_1\hat{a}_2e^{-2i\tilde{\omega}_0 t} - \hat{a}_2^\dagger\hat{a}_1 -\hat{a}_1^\dagger\hat{a}_2)\\
+ \sum_{j=1,2}i \left( \hat{a}_j(e^{-i(\tilde{\omega}_0+\omega_j)t- i\phi  }+ e^{-i(\tilde{\omega}_0-\omega_j)t + i\phi } ) -  \hat{a}_j^{\dagger }(e^{i(\tilde{\omega}_0+\omega_j)t + i\phi  }+ e^{i(\tilde{\omega}_0-\omega_j)t - i\phi } )  \right)f_j(t)\;,
\end{multline}
where $ \delta_j= \bar{ \omega}_0-\omega_0 $ is the detuning of each qubit  from the initial harmonic frequency. The simulation time of the gmon system $ t $ is usually set to be much longer than $ 1/\eta $ satisfying  $ t \tilde{\omega}_0 \gg1 $.  We can therefore apply the rotating-wave-approximatoin~(RWA) to omit   the highly oscillating components with  phase oscillating equal or faster than $ \tilde{\omega}_0  $. As a result, we obtain the Hamiltonian for the gmon circuit bosonic modes in RWA basis as
\begin{multline}
\hat{H}_{\RWA}=   \frac{\eta}{2} \sum_{j=1}^2 \hat{n}_j(  \hat{n}_j- 1  ) +g(t) ( \hat{a}_2^\dagger\hat{a}_1+\hat{a}_1^\dagger\hat{a}_2)  + \sum_{j=1}^2\delta_j (t) \hat{n}_j +\sum_{j=1}^2i f_j(t)\left( \hat{a}_j e^{    i\phi_j(t) } -  \hat{a}_j^{\dagger } e^{-  i\phi_j(t) }  \right)\;,
\end{multline}
where we assumed that the microwave frequencies $ \omega_j $ are not too far from the qubit frequency $ \tilde{\omega}_0 $  such that $ \Delta \omega= \tilde{\omega}_0 -\omega_j  \ll\tilde{\omega}_0 $, and define    $ g(t)=\alpha^\prime(t)\frac{\kappa_1\kappa_2}{2}  $.

\begin{table}[h]
\centering
\begin{tabular}{c @{\hspace{0.3cm}} c@{\hspace{0.3cm}} c@{\hspace{0.3cm}}c@{\hspace{0.3cm}}c@{\hspace{0.3cm}}c@{\hspace{0.3cm}}c } \hline \hline
  & $\eta $ &   $g(t)$ & $\delta_j(t)$  & $ f_j(t) $ &  $ \phi_j(t) $  \\\hline\hline
amplitude  &200 MHz & [-20, 20 ] MHz& [-20, 20 ] MHz  &  [-20, 20 ] MHz&   [0, 2$\pi$] \\\hline
error amplitude & $\pm$ 1 MHz &$\pm$ 1 MHz&$\pm$1 MHz& $\pm$ 1 MHz&     \\
\hline \hline
\end{tabular} 
\caption{Hamiltonian control parameter range.} \label{Table1}
\end{table}
The above Hamiltonian describes the two coupled Josephson L-C circuits in the perturbative regime of the anharmonicity. The parameter range of its coupling terms are listed in Table.~\ref{Table1}.
We require that the magnitudes  of  our two-qubit coupling g-pulse $ g(t) $, detuning $ \delta_j(t) $ and microwave pulse amplitude $ f_j(t) $ are at least one magnitude lower than $ \eta $  to meet our   leakage minimization condition. Moreover, we make the modulation rate of all  parameters $ g(t), \delta_j(t), f_j(t) $ to be within the frequency bandwidth $ [-50, 50 ] $ MHz  due to experimental limitations. Projected onto the qubit basis, this Hamiltonian takes the form:
\begin{align}
\hat{H}_{\RWA}= \frac{g(t)}{2}(\sigma_1^x\sigma_2^x + \sigma_1^y\sigma_2^y ) \sum_{j=1}^2 \left[ \frac{\delta_j(t)}{2}\sigma_j^z  - f_j(t) \left(\sin\phi_j(t) \sigma_j^x + \cos\phi_j(t) \sigma_j^y \right)   \right].
\end{align}

The stochastic training environment is implemented by adding amplitude fluctuation   sampled from a  zero mean Gaussian distribution of $ 1 $ MHz variance to the six control amplitudes $ \eta \to \eta + \delta\eta $, $ g(t_k)\to g(t_k) + \delta g(t_k) $, $ \delta_j(t_k) \to \delta_j(t_k) + \delta \delta_j(t_k) $ and $ f_j(t_k) \to \delta f_j(t_k) $ in every discretized time step $ t_k \in [ 0, \Delta T, 2\Delta T, \ldots, N \Delta T]$.

\section{Complete Leakage Bound}\label{app:clb}

In the following sections, we focus on deriving the correct form of    $ L_{\text{tot}} $ to fully account for differents sources of leakage errors  during a time-dependent Hamiltonian evolution. We start with the formulation of TSWT for defining a rotated computational basis, where direct coupling induced leakage errors are suppressed to the higher order.  Next, we prove that  the non-adiabatic leakage  from our generalization of adiabatic theorem under TSWT is sub-dominant  to obtain the final  leakage bound  $ L_{\text{tot}} $ defined  in Eq.~(9) of the main text.

\subsection{Time-dependent Schrieffer-Wolff transformation}\label{TSW}
The success of Schrieffer-Wolff transformation~\cite{Goldin2000} rests upon the difference in energy scales  between the energy gap separating different subspaces and  the  coupling terms within and between different  subspaces.    To generalize the previous formulation, we decompose  the system Hamiltonian  into three parts 
\begin{align}\label{totalTSWHamiltonian}
& \hat{H}(t)=\hat{H}_0+\hat{H}_1(t) +\hat{H}_2(t),\\\label{H0}
 & \hat{H}_0 =\sum_\alpha\sum_{m \in\Omega_\alpha} E_1\ket{m}\bra{m}, \\\nonumber
 &\hat{H}_1(t) =\sum_\alpha \sum_{m \in \Omega_\alpha}\bra{ m}\hat{H}_1^{\alpha }(t)\ket{m^\prime}\ket{m}\bra{m^\prime}\\\nonumber
& \hat{H}_2(t)= \sum_{\alpha\neq \alpha^\prime}\sum_{m \in \Omega_\alpha, ,m^\prime\in \Omega_{\alpha^\prime}}\bra{ m}\hat{H}_2^{\alpha,\alpha^\prime}(t)\ket{m^\prime} \ket{\alpha, m}\bra{\alpha^\prime, m^\prime}\;,
\end{align} 
where     $ \Omega_\alpha $ with $ \alpha\in\{0,1,2,\ldots \} $ represents different subspaces  and $ \ket{m} $ denotes orthogonal basis state spanning each same subspace, the first  term $ \hat{H}_0 $ accounts for   the time-independent part of  the system Hamiltonian, the   second term $  \hat{H}_1(t)$ accounts for the time-dependent coupling within each subspace $ \Omega_\alpha $ which we call the ``block-diagonal term", and the third part  $  \hat{H}_2 (t)$ accounts for the coupling between different subspaces which we call the `` block-off-diagonal term". In order to apply the perturbative expansion of TSWT, we assume the magnitudes of three parts of the system Hamiltonian  obey: $ \Delta= \min_{ \alpha \neq 0}|E_\alpha-E_0| \gg |\bra{ m}\hat{H}_1^{\alpha }(t)\ket{m^\prime}|  \sim|\bra{ m}\hat{H}_2^{\alpha,\alpha^\prime}(t)\ket{l}|~\epsilon , \, \text{for} \,\, \vee m, m^\prime \in \Omega_\alpha, l \in \Omega_{\alpha^\prime} , \alpha\neq \alpha^\prime  $ through out the time-dependent Hamiltonian evolution, where we use $ \Delta $ and $ \epsilon $ to represent the different energy scales satisfying $ \epsilon/\Delta  <<1 $.  Under this assumption, each subspace $ \Omega_\alpha $ is separated from the other  by   energy gap much larger than the coupling within $ \Omega_\alpha $ or between $ \Omega_\alpha $ and $ \Omega_\beta $ for $ \beta \neq \alpha $.   Now we perform the Schrieffer-Wollf  transformation  to  rotate  the original basis  state $ \ket{\psi} $ to  $ \ket{\tilde{\psi }}=e^{-\hat{S}}\ket{\psi}$. In this rotated basis,   the effective Hamiltonian $  \hat{\mathbb{H}} $ can be found by
\begin{align}  
i\frac{d}{dt}\ket{\tilde{\psi }}&=\hat{\mathbb{H}} \ket{\tilde{\psi}}=i \frac{d e^{-\hat{S}}}{dt } e^{\hat{S}}\ket{\tilde{\psi }} + ie^{-\hat{S}}\frac{d}{dt}\ket{\psi},\\ \label{effectiveHam2}
 \hat{\mathbb{H}}&=-i \sum_{j=0}^\infty \frac{1}{(j+1)!}[\dot{\hat{S}}, \hat{S}]_j  + e^{-\hat{S}}\hat{H} e^{\hat{S}}\;,
\end{align}
where the anti-Hermitian operator $ \hat{S}(t) $  contains  non-zero term only between different subspaces and is thus block-off-diagonal.  The goal of TSWT is to find perturbative solution  of the rotation $ \hat{S}(t)= \epsilon \hat{S}_1(t) + \epsilon^2 \hat{S}_2(t) + \ldots  +\epsilon^n \hat{S}_n(t)  $ that   block-diagonalizes the system Hamiltonian~\cite{Goldin2000} such that the effective Hamiltonian $ \hat{\mathbb{H}} $'s block-off-diagonal terms  are suppressed to an order of $ O\left( \frac{\epsilon^{n+1} }{\Delta^n} \right)$ for $ n $th order perturbative solution of $\hat{S}(t)   $.  We provide the derivation  of $ \hat{S}(t) $ up to the second order as a function of $ \hat{H}(t) $  below.

We expand the expression for the effective Hamiltonian in Eq.~(\ref{effectiveHam2}) into 
\begin{align}  \label{effectiveH0}
 \hat{\mathbb{H}}&=-i \sum_{j=0}^\infty \frac{1}{(j+1)!}[\dot{\hat{S}}, \hat{S}]_j  +   \sum_{j=0} \frac{1}{j ! }[\hat{H}(t), \hat{S}(t)]^{j}
\end{align}
using the algebraic relation 
\begin{align}\label{SWHamiltonian}
e^{-\hat{S}(t)}\hat{H}(t)e^{\hat{S}(t)}= \sum_{j=0} \frac{1}{j ! }[\hat{H}(t), \hat{S}(t)]^{j},\\\label{commutationEq}
[\hat{H}(t), \hat{S}(t)]^{j}=[\ldots [ [ \hat{H}(t),  \underbrace{\hat{S}(t)] , \hat{S}(t)]\ldots ,  \hat{S}(t)]}_{\text{j  many }}.
\end{align} 
We can now separate Eq.~(\ref{effectiveH0}) into block-diagonal and block-off-diagonal component and choose correct form of   rotation $ \hat{S}(t) $ to cancel  the block-off-diagonal component to a given order. For simplicity, henceforth we do not write down the time-dependency $ (t) $ for each operator. 

As defined in the main text, $ \hat{H}_0 $ and $ \hat{H}_1 $ are block-diagonal, while $ \hat{S} $  and $ \hat{H}_2 $ are block-off-diagonal.  Therefore, even orders of the commutation in Eq.~(\ref{commutationEq}) between  $ \hat{H}_0 $ or $ \hat{H}_1 $ with $ \hat{S} $  or  $\dot{ \hat{S}} $   are block-diagonal, while their odd  counterparts are block-off-diagonal. And the odd orders of commutations between $ \hat{H}_2 $  or  $\dot{ \hat{S}} $ with  $ \hat{S} $ are block-diagonal and their odd counterparts are block-off-diagonal. We  thus obtain   block-diagonal part  of  Eq.~(\ref{SWHamiltonian}):
\begin{align}\label{diagH}
\hat{\mathbb{H}}_{\dd}= -i \sum_{j=0}^\infty \frac{1}{(2j+2)!}[\dot{\hat{S}}, \hat{S}]^{2j+1} +  \sum_{j=0}^\infty\frac{1}{2j !}[\hat{H}_0+ \hat{H}_1, \hat{S}]^{2j}+\sum_{j=0}^\infty\frac{1}{(2j +1)!}[\hat{H}_2, \hat{S}]^{2j+1}.
\end{align}
Similarly we obtain the block-off-diagonal part  of the effective Hamiltonian:
\begin{align}\label{offdiagH}
\hat{\mathbb{H}}_{\od}= -i \sum_{j=0}^\infty \frac{1}{(2j+1)!}[\dot{\hat{S}}, \hat{S}]^{2j } +  \sum_{j=0}^\infty\frac{1}{(2j+1) !}[\hat{H}_0+ \hat{H}_1, \hat{S}]^{2j+1}+\sum_{j=0}^\infty\frac{1}{ 2j !}[\hat{H}_2, \hat{S}]^{2j}.
\end{align}

Now we like to set the block-off-diagonal component to zero up to the order $ O(\epsilon^3/\Delta^2) $.  With a slight abuse of notation, we use below $ \epsilon $ in short for the unitless value $ \epsilon/\Delta $. We  solve such diagonalization through  perturbative expansion of the TSWT rotation:    $ \hat{S} = \epsilon \hat{S}_1  + \epsilon^2 \hat{S}_2 + \ldots  +\epsilon^n \hat{S}_n   $ and   $\dot{ \hat{S}} = \epsilon^2\dot{ \hat{S}}_1  + \epsilon^3 \dot{\hat{S}}_2 + \ldots  +\epsilon^{n+1}\dot{ \hat{S}}_n   $   such that we solve   order  by order   rotation $\hat{S}_n(t)   $  that cancels the  block-off-diagonal parts of the Hamiltonian of order $ O(\epsilon^{n} )$. Here, we adopt the convention that the time derivative of each order of rotation is one order higher: $| \dot{ \hat{S}}_n| \sim \epsilon|\hat{S}|   $.

Following the same convention, we rewrite the original Hamiltonian as $ \hat{H} =\hat{H}_0+\epsilon(\hat{H}_1  +\hat{H}_2 ) $ according to  relative amplitudes of different components, and insert it together with expansion of $ \hat{S}(t) $  to obtain the order by order perturbative expansion of block-diagonal and block-off-diagonal parts of the effective Hamiltonian:
\begin{align}\label{diagonalH}
\hat{\mathbb{H}}_{\dd}=& \hat{H}_0 + \epsilon \hat{H}_1 +\frac{1}{2}\epsilon^2 [\hat{H}_2, \hat{S}_1] + \frac{1}{2}\epsilon^3[\hat{H}_2, \hat{S}_2] - i \frac{1}{2}\epsilon^3[\dot{\hat{S}}_1,\hat{S}_1]+ O\left( \epsilon^4 \right),\\\label{offdiagonalH}
\hat{\mathbb{H}}_{\od}=& \epsilon [\hat{H}_0, \hat{S}_1] +  \epsilon \hat{H}_2 +  \epsilon^2[\hat{H}_0,\hat{S}_2]+\epsilon^2[\hat{H}_1,\hat{S}_1] -i \epsilon^2\dot{\hat{S}}_1 \\\nonumber
&+   \epsilon^3 [\hat{H}_1, \hat{S}_2] +\frac{\epsilon^3 }{3}[[\hat{H}_2, \hat{S}_1],\hat{S}_1] -i \epsilon^3\dot{\hat{S}}_2+ O\left(\epsilon^4\right).
\end{align}
Perturbatively diagonalizing the Hamiltonian, to the first order in $ \epsilon $ we have  
\begin{align}
[\hat{H}_0, \hat{S}_1]+ \hat{H}_2 =0\;,
\end{align}
which gives us the matrix expression for the first order SW rotation
\begin{align}\label{S1eq}
\hat{S}_1^{\alpha, \alpha^\prime}=\frac{\hat{H}_2^{\alpha, \alpha^\prime}}{E_{\alpha^\prime}-E_\alpha}
\end{align}
where $ \hat{H}_2^{\alpha, \alpha^\prime} $ is  the Hamiltonian between subspace $ \Omega_\alpha $ and $ \Omega_{\alpha^\prime} $ and is itself a matrix. To the second order in diagonalization, we have
\begin{align}
[\hat{H}_0,\hat{S}_2]+[\hat{H}_1,\hat{S}_1]- i \dot{\hat{S}}_1=0\;,
\end{align}
which immediately yields the matrix representation of the second order SW rotation between subspace $ \Omega_\alpha $ and $ \Omega_\alpha^\prime $:
\begin{align}\label{S2eq}
\hat{S}_2^{\alpha, \alpha^\prime}= \frac{\hat{H}_1^{\alpha}\hat{H}_2^{\alpha, \alpha^\prime}-\hat{H}_2^{ \alpha, \alpha^\prime}\hat{H}_1^{\alpha^\prime}}{(E_{\alpha^\prime}-E_\alpha)^2} -\frac{i\dot{ \hat{H}}_2^{\alpha, \alpha^\prime}}{(E_{\alpha^\prime}-E_\alpha)^2}
\end{align}
where we use $ \hat{H}_1^\alpha $ as the sub-dominant Hamiltonian terms within the subspace $ \Omega_\alpha $.  Inserting Eq.~(\ref{S1eq}) and (\ref{S2eq}) into Eq.~(\ref{diagonalH}) and (\ref{offdiagonalH}) gives us the block-diagonal and block-off-diagonal parts of the effective Hamiltonian after the second   order TSWT:
\begin{align} 
(\hat{\mathbb{H}}_{\dd})^{\alpha}
=&\hat{H}_0^\alpha +    \hat{H}_1^\alpha - \sum_{\alpha^\prime\neq \alpha}\frac{\hat{H}_2^{\alpha, \alpha^\prime} \hat{H}_2^{\alpha^\prime, \alpha} }{(E_{\alpha^\prime} -E_\alpha)}  \\\nonumber
&+ \frac{1}{2}\sum_{\alpha^\prime \neq \alpha} \frac{\hat{H}_2^{\alpha,\alpha^\prime}\hat{H}_2^{\alpha^\prime, \alpha}\hat{H}_1^{\alpha}- \hat{H}_1^{\alpha}\hat{H}_2^{\alpha,\alpha^\prime}\hat{H}_2^{\alpha^\prime, \alpha}}{(E_{\alpha^\prime} -E_\alpha)^2}\\\nonumber
&+ \sum_{\gamma\neq \alpha} \frac{i\left[ \dot{\hat{H}}_2^{\alpha,\gamma}\hat{H}_2^{\gamma, \alpha}- \hat{H}_2^{ \alpha,\gamma} \dot{\hat{H}}_2^{ \gamma,\alpha} \right]}{ (E_\alpha-E_\gamma)^2} + O(\epsilon^4),\\\nonumber 
(\hat{\mathbb{H}}_{\od})^{\alpha,\alpha^\prime}&=\hat{H}_{\od}^{\alpha,\alpha^\prime} +\Delta \hat{H}_{\od}^{\alpha,\alpha^\prime},\\\nonumber
= &\frac{ (\hat{H}_1^\alpha)^2 \hat{H}_2^{\alpha,\alpha^\prime}-2\hat{H}_1^\alpha\hat{H}_2^{\alpha,\alpha^\prime}\hat{H}_1^{\alpha^\prime}+ \hat{H}_2^{\alpha,\alpha^\prime} (\hat{H}_1^{\alpha^\prime})^2 }{(E_{\alpha^\prime}-E_\alpha)^2} 
\\\nonumber
&+  \frac{2}{3(E_{\alpha^\prime}-E_\alpha)}\sum_\gamma \left[\frac{ \hat{H}_2^{\alpha, \alpha^\prime} \hat{H}_2^{ \alpha^\prime, \gamma} \hat{H}_2^{\gamma, \alpha^\prime}}{E_{\alpha^\prime}-E_\gamma} - \frac{\hat{H}_2^{\alpha,\gamma}\hat{H}_2^{\gamma,\alpha} \hat{H}_2^{\alpha, \alpha^\prime}}{E_\alpha-E_\gamma} \right] \\\label{HEFFoff}
&-i \left[ \frac{\dot{\hat{H}}_1^{\alpha}\hat{H}_2^{\alpha, \alpha^\prime}+ 2\hat{H}_1^{\alpha}\dot{\hat{H}}_2^{\alpha, \alpha^\prime}-2\dot{\hat{H}}_2^{ \alpha, \alpha^\prime}\hat{H}_1^{\alpha^\prime}-\hat{H}_2^{ \alpha , \alpha^\prime}\dot{\hat{H}}_1^{\alpha^\prime}}{(E_{\alpha^\prime}-E_\alpha)^2} \right]  -\frac{ \ddot{ \hat{H}}_2^{\alpha, \alpha^\prime}}{(E_{\alpha^\prime}-E_\alpha)^2}+ O(\epsilon^4)
\end{align}
With this block-off-diagonal Hamiltonian in hand, we continue to evaluate the   population leakage out of the qubit subspace caused by this direct coupling in both   off-resonant    and on-resonant regimes in the following section.

\subsection{Leakage  Bound}\label{LeakageBoundSec}

\noindent\textit{Direct Coupling Leakage   Bound}:

The non-zero block-off-diagonal part of the Hamiltonian $ \hat{\mathbb{H}}_{\od} $  directly couples the qubit subspace to the higher energy subspace. To evaluate the population that transition out of the qubit subspace due to this direct couplings, we adopt the interacting picture with state basis   $ \ket{ \psi(t)}_I =U_{\dd}^{-1}(t)\ket{ \psi(t)} $  that relates to the Schr\"{o}dinger picture basis initial state $ \ket{ \psi(0)} $ by a  block-diagonal Hamiltonian evolution  $U_{\dd}(t)=\mathbb{T} [ e^{-i \int_0^t \hat{\mathbb{H}}_{\dd}(\tau) d\tau}]  $, where $ \mathbb{T}  $ represents the time-ordering. The Schr\"{o}dinger equation in the interacting picture is $ i\frac{d}{dt}\ket{ \psi(t)}_I  = U_{\dd}^{-1}(t)\hat{\mathbb{H}}_{\od}(t) U_{\dd}(t) \ket{ \psi(0)}_I  $ with the lowest order solution given $ |\hat{\mathbb{H}}_{\od}(t)|  $ is of order $  O( \epsilon^3/\Delta^2)  $:
\begin{align}\label{interactPsi} 
&\ket{ \psi(t)}_I \approx\left[   I -i \int_{0}^t U_{\dd}^{-1}(\tau)\hat{\mathbb{H}}_{\od}(\tau) U_{\dd}(\tau) d\tau \right]\ket{ \psi(0)} 
\end{align}
where we insert  the initial condition $\ket{ \psi(0)}_I = \ket{ \psi(0)}$.  Since $ U_{\dd}(t) $ preserve the computational subspace, the   leakage is thus  evaluated by the  sum of the amplitudes of all excited states outside the qubit subspace  due to the  non-zero block-off-diagonal Hamiltonian after TSWT:
\begin{align}\nonumber
&L_{\text{direct}}(t)=\sum_{\alpha\neq 0, m\in \Omega_\alpha}\left|\bra{  m (t)} \left( U_{\dd}(0) \ket{ \psi(0)}-U_{\dd}(t)  \ket{ \psi(t) }_I \right)\right|  \\\nonumber
&=\sum_{\alpha\neq 0, m\in \Omega_\alpha}\left|\bra{  m(t)}    \int_{0}^t U_{\dd} (t, \tau)\hat{\mathbb{H}}_{\od}(\tau) U_{\dd}(\tau,t) d\tau  \ket{ \psi(t)}\right| \\\nonumber
&\stackrel{1}{\approx}\sum_{\alpha\neq 0, m\in \Omega_\alpha}\left|\bra{  m(t)}  \left. \left(    U_{\dd} (t, \tau)  \frac{1}{\Delta_\alpha}\hat{\mathbb{H}}_{\od}(\tau) U_{\dd}(\tau,t) d\tau   \right)\right|_{\tau=0}^t\right. \\\nonumber
&-  \left.   \int_0^t U_{\dd} (t, \tau)  \frac{1}{\Delta_\alpha}\frac{d}{d\tau} \left( \hat{\mathbb{H}}_{\od}(\tau)\hat{\mathbb{H}}_{\dd}^{-1} \frac{d  U_{\dd}(\tau,t)}{d\tau}\right) d\tau  \ket{ \psi(t)}\right|\\\nonumber
&\stackrel{2}{=}\sum_{\alpha\neq 0, m\in \Omega_\alpha}\left|\bra{  m(t)}    \left. \left(    U_{\dd} (t, \tau)  \frac{1}{\Delta_\alpha}\hat{\mathbb{H}}_{\od}(\tau) U_{\dd}(\tau,t) d\tau   \right) \right|_{\tau=0}^t\right.\\\nonumber
& - \left. \left.  \left(  2U_{\dd} (t, \tau)  \frac{1}{\Delta_\alpha^2}\frac{d \hat{\mathbb{H}}_{\od}(\tau) }{d\tau}    \frac{d  U_{\dd}(\tau,t)}{d\tau}   \right)\right|_{\tau=0}^t + \int_0^tU_{\dd} (t, \tau)  \frac{1}{\Delta_\alpha^2}\frac{d^2 \hat{\mathbb{H}}_{\od}(\tau) }{d\tau^2} U_{\dd}(\tau,t)  \ket{ \psi(t)}\right|\\\label{finalBounddirectCoupling}
&\stackrel{3}{ \leq  }   \frac{ \Vert \hat{\mathbb{H}}_{\od}(0) \Vert }{\Delta(0)}+ \frac{ \Vert \hat{\mathbb{H}}_{\od}(t) \Vert }{\Delta(t)} +   2\frac{ \Vert \dot{ \hat{\mathbb{H}}}_{\od}(0) \Vert }{\Delta^2(0)}+ 2\frac{ \Vert \dot{\hat{\mathbb{H}}}_{\od}(t) \Vert }{\Delta^2(t)} +  \int_0^t \frac{1}{\Delta^2(t)} \left|\left| \frac{d^2 \hat{\mathbb{H}}_{\od}(t)}{dt^2} \right|\right|dt
\end{align}
where   the approximation 1 is obtained by the integration by part , replacing $ U_{\dd}( \tau, t) $ with $ \frac{dU_{\dd}( \tau,t)}{d\tau}   \hat{\mathbb{H}}_{\dd}^{-1}$ and that $ \bra{m_\alpha} \hat{\mathbb{H}}_{\dd}^{-1}\ket{\psi_0}\sim 1/\Delta_\alpha $; the equality 2 is obtained by two more integration by parts for the last integration from the previous line, and the inequality 3 is given by the triangle inequality.
The middle two terms of Eq.~(\ref{finalBounddirectCoupling}) are shown~(see SM.~\ref{LeakageBoundSec})  to be at least one magnitude  smaller than the  left of the terms  in either on-resonant or off-resonant regime  and can thus be omited in our   final leakage bound:
\begin{align}\label{apptotalLeakageBound}
L_{tot}= \frac{ \Vert \hat{\mathbb{H}}_{\od}(0) \Vert }{\Delta(0)}+ \frac{ \Vert \hat{\mathbb{H}}_{\od}(t) \Vert }{\Delta(t)} +  \int_0^t \frac{1}{\Delta^2(t)} \left|\left| \frac{d^2 \hat{\mathbb{H}}_{\od}(t)}{dt^2} \right|\right|dt
\end{align}

We  now compare different terms of the leakage bound of Eq.~(9) in the main text  in  both on-resonant and off-resonant frequency regimes separately.   First, consider  the off-resonant frequency regime such that the Fourier components of the Hamiltonian modulation has a frequency range upper bounded by a value much smaller than  $\Delta $ such that for any $ n $th order  time-derivative  we have  $  \frac{d^n  \vert\vert \hat{\mathbb{H}}_{\od}\vert\vert /dt^n}{    \vert \vert \hat{\mathbb{H}}_{\od}  \vert \vert  } \sim \epsilon^n \ll \Delta^n$, where the modulation rate of the system Hamiltonian is much smaller than the minimum energy gap separating different energy subspaces. The    TSWT assumption in off-resonant regimes  implies: $  \Vert \hat{\mathbb{H}}_{\dd}\Vert \sim \epsilon $ and $ \Vert   \hat{\mathbb{H}}_{\od} \Vert \sim \epsilon^3/\Delta^2 $.  Moreover, in the off-resonant regime, the total run time $ T $ obeys $1/T  \ll  \Delta   $ to gaurantee the modulation frequency to be smaller than the energy gap.  The first two terms in Eq.~(\ref{finalBounddirectCoupling})  $  \frac{ \Vert \hat{\mathbb{H}}_{\od}(0) \Vert }{\Delta(0)}+ \frac{ \Vert \hat{\mathbb{H}}_{\od}(t) \Vert }{\Delta(t)} \sim O(\frac{\epsilon^3}{\Delta^3} )$ dominate in this regime since they separated from the left of the terms by a factor of  $1/\Delta  $ with  $\Delta =\min_t \Delta(t)  $.

Second, consider the on-resonant frequency regime where the frequency components of the Hamiltonian modulation is of similar magnitudes as the energy gap such that: $  \frac{d^n  \vert\vert \hat{\mathbb{H}}_{\od}\vert\vert /dt^n}{    \vert \vert \hat{\mathbb{H}}_{\od}  \vert \vert  } \sim \Delta^n$,  where   unwanted couplings between the qubit subspace and a higher energy subspace is of approximately the same frequency  as the energy gap. TSWT assumption still holds with the amplitudes of the time-dependent Hamiltonians obeying $ \Vert \hat{H}_{1}\Vert  \sim \Vert \hat{H}_{2}\Vert \sim \epsilon $.  The last term  in Eq.~(\ref{finalBounddirectCoupling})  $\int_0^t \frac{1}{\Delta^2(t)} \left|\left| \frac{d^2 \hat{\mathbb{H}}_{\od}(t)}{dt^2} \right|\right|dt \sim O(\frac{\epsilon^3}{\Delta^3}) $ dominate in this regime since they separated from the left of the terms  by a factor of $1/\Delta  $.

\noindent\textit{Non-adiabatic Leakage   Bound}:

A generalized adiabatic theorem below bounds the   leakage amplitude into higher energy subspaces.

\noindent\textbf{Theorem 2.} Let $ \hat{H}(s) $ be a twice differentiable Hamiltonian  parametrized by a unit-free re-scaled time $ s\in[0,1] $  comprising three parts: $ \hat{H}(s)=H_0+ \hat{\mathbb{H}}_{\dd}(s) + \hat{\mathbb{H}}_{\od}(s)  $.  The time-invariant term
$   \hat{H}_0 =\sum_{\alpha=0}^\infty\sum_{m \in\Omega_\alpha} E_\alpha\ket{m}\bra{m}$ ensures large constant energy gap between  the lowest energy subspace $ \Omega_0 $ and other higher energy subspaces. The time-varying term  $ \hat{\mathbb{H}}_{\dd}(s)  $ accounts  for couplings within each non-degenerate subspace $ \Omega_\alpha $ and  $  \hat{\mathbb{H}}_{\od}(s) $  accounts for the coupling between   different subspaces.  There is a separation between the energy gap and  inter/intra-subspace coupling: $ \Delta = \min_{\alpha}|E_\alpha-E_0|\gg | \hat{\mathbb{H}}_{\dd}(s)|\sim | \hat{\mathbb{H}}_{\od}(s) |\sim \delta$ . Let  $ \ket{\phi_0(s )}=\sum_{m\in \Omega_0} a_m(s)\ket{m} $ be an instantaneous eigenstate in the lowest energy subspace $ \Omega_0 $ at physical time $sT $.  Let $ \ket{\psi(s )} $ be the state evolved from the same initial state $ \ket{\phi_0(0)}$ at time $ s=0 $ under  the total Hamiltonian $ \hat{H}(s ) $ to time $ s $.  We have the following inequality that bound the difference between these two states at the final time $ T $ by:
\begin{align}\label{nonadiabaticLeakageBound}
&L_{\text{non-adiabatic}} \leq\frac{1}{T}  \left[ \frac{1}{\Delta^2(s) } ( ||\frac{d \hat{\mathbb{H}}_{\od}(s)}{ds} || + T||[ \hat{\mathbb{H}}_{\dd}(s),\hat{\mathbb{H}}_{\od}(s)]||)_{s=1} +  \frac{1}{\Delta^2(s) }   ( ||\frac{d \hat{\mathbb{H}}_{\od}(s)}{ds} || \right. \\\nonumber
& \left. +T ||[ \hat{\mathbb{H}}_{\dd}(s), \hat{\mathbb{H}}_{\od}(s)]||)_{s=0} \right] + \int_0^1 \frac{5}{\Delta^3(s)} \left(  ||\frac{d \hat{\mathbb{H}}_{\od}(s)}{ds} || +T ||[ \hat{\mathbb{H}}_{\dd}(s), \hat{\mathbb{H}}_{\od}(s)]|| \right)^2 ds    \\\nonumber
&+ \int_0^1 \frac{1}{\Delta^2(s)} \left(T ||[\hat{\mathbb{H}}_{\dd}(s), [\hat{\mathbb{H}}_{\dd}(s), \hat{\mathbb{H}}_{\od}(s)]] || + 2|| [\hat{\mathbb{H}}_{\dd}(s),\frac{d\hat{\mathbb{H}}_{\od}(s)}{ds} ]|| + 2 ||[ \frac{d \hat{\mathbb{H}}_{\dd}(s) }{ds} ,\hat{\mathbb{H}}_{\od}(s) ]||\right. \\\nonumber
&\left. +\frac{1}{T}|| \frac{d^2 \hat{\mathbb{H}}_{\od}(s)}{ds^2} || \right) ds\;,
\end{align} 
 where we choose appropriate global phase for the initial state following the convention in \cite{Jordan2008}. 
 
\noindent\textit{Proof}: The proof is a generalization of Goldstone in Ref.~\cite{Jordan2008} to account for inter-subspace dynamics induced by $  \hat{\mathbb{H}}_{\dd} $ and intra-subspace coupling induced by $ \hat{\mathbb{H}}_{\od} $. 
To begin with, we change from Schr\"{o}dinger picture basis $ \ket{  \psi(s)}  $ to the interaction picture basis $ \ket{\tilde{ \psi}(s)}  $.   Let us  define the unitary evolution under the time-dependent diagonal Hamiltonian $  \hat{\mathbb{H}}_{\dd}(s) $ as  
\begin{align}\label{diagonalU}
U_{\dd}(s,0)= \mathbb{T} [e^{-iT\int_0^1   \hat{\mathbb{H}}_{\dd}(s) ds}]
\end{align}
which includes quantum dynamics  within each subspace $ \Omega_\alpha $ and satisfies the Schr\"{o}dinger's equation:
\begin{align}
 \frac{d}{ds }U_{\dd}(s,0)=-iT(  \hat{\mathbb{H}}_{\dd}(s) )U_{\dd}(s,0).
\end{align}
In the rotated basis $ \ket{\tilde{ \psi}(s)} =U_{\dd}(s,0)\ket{  \psi(s)}  $,  the quantum dynamics induced by $  \hat{\mathbb{H}}_{\dd}(s)  $ within each subspace is absent since:
\begin{align}\label{basistransform1}
&\frac{i}{T}\frac{d}{ds}\ket{\psi(s)}=(\hat{H}_0+ \hat{\mathbb{H}}_{\dd}(s) +\hat{\mathbb{H}}_{\od}(s))U_{\dd}(s,0)\ket{\tilde{ \psi}(s)},\\
&\frac{i}{T}\frac{d}{ds}\left(U_{\dd}(s,0)\right)\ket{\tilde{ \psi}(s)} + U_{\dd}(s,0)\frac{d}{ds}\ket{\tilde{ \psi}(s)}=(\hat{H}_0+ \hat{\mathbb{H}}_{\dd}(s) +\hat{\mathbb{H}}_{\od}(s))U_{\dd}(s,0)\ket{\tilde{ \psi}(s)},\\
\to & \frac{d}{ds}\ket{\tilde{ \psi}(s)}=U_{\dd}(0,s)(\hat{H}_0 +\hat{\mathbb{H}}_{\od}(s))U_{\dd}(s,0)\ket{\tilde{ \psi}(s)}=\hat{\tilde{H}}(s) \ket{\tilde{ \psi}(s)},\\
& \hat{\tilde{H}}(s)=U_{\dd}(0,s)(\hat{H}_0 +\hat{\mathbb{H}}_{\od}(s))U_{\dd}(s,0)=\hat{H}_0 + U_{\dd}(0,s)\hat{\mathbb{H}}_{\od}(s))U_{\dd}(s,0)
\end{align}
where we use   $ \hat{\tilde{H}}(s) $ to represent the effective Hamiltonian in the rotated basis.  Now we continue to establish the adiabatic theorem in the rotated basis: if the system Hamiltonian changes slow enough, the transition from the ground state living in the lowest energy subspace $ \Omega_0 $ to an excited state in higher energy subspace is negligible.

Let $ \ket{\tilde{\phi}_n(s)} $ be the instantaneous energy eigenstate of the effective Hamiltonian in the interaction picture :
\begin{align}\label{instantaneousBasis}
\hat{\tilde{H}}(s)  \ket{\tilde{\phi}_n(s)}=\tilde{E}_n(s) \ket{\tilde{\phi}_n}.
\end{align}
The caveat is here the eigenvalues $ \{\tilde{E}_n\} $ are that of the effective Hamiltonian $ \hat{H}_0 + U_{\dd}(0,s)\hat{\mathbb{H}}_{\od}(s))U_{\dd}(s,0) $, which differs from the original energies $ \{ E_n \} $ of the static Hamiltonian $ \hat{H}_0=\sum_{n=0}^\infty E_n \ket{\phi_n}\bra{\phi_n} $ by a factor  of    $  O( \frac{\epsilon^3}{\Delta^3}) $ under our  assumptions.

By appropriately choosing the global phase $ e^{i\int_0^T E_0(t) dt} $ of initial state, we make sure that the lowest energy state in the subspace $ \Omega_0 $ has an eigenvalue $ E_0=0 $ and will thus remain zero through out the evolution:
\begin{align}\label{groundstate}
\hat{\tilde{H}}(s)  \ket{\tilde{\phi}_0(s)}=0.
\end{align}
The deviation from the adiabatic evolution between the actual state $  $ and the instantaneous eigenstate is    measured by 
\begin{align}\label{deviation}
\delta(T )= \ket{\tilde{\psi}(T)}- \ket{\tilde{\phi}_0(T)}=U_{T}(1,0)\ket{\tilde{\phi}_0(0)}- \ket{\tilde{\phi}_0(T)}\;,
\end{align}
where we use $ U_{T}(1,0) $ as  the unitary transformation induced by   the effective Hamiltonian in the rotated basis $ \hat{\tilde{H}}(s) $ for time $ T $:
\begin{align}
U_{T}(1,s)=\mathbb{T} [e^{-iT\int_s^1\hat{\tilde{H}}(s) ds}]\;,
\end{align}
which also obeys the Schr\"{o}dinger's equation:
\begin{align}\label{SchrodingerUT}
 \frac{d}{ds }U_{T}(1,s)=iT  U_{T}(1,s)\hat{\tilde{H}}(s) = iT U_{T}(1,s)( \hat{H}_0 + U_{\dd}(0,s)\hat{\mathbb{H}}_{\od}(s))U_{\dd}(s,0)).
\end{align}
Similar to proof in Goldstone\cite{Jordan2008}, we re-write the adiabatic deviation in Eq.~(\ref{deviation}) as
\begin{align}
\delta(T)= \int_0^1 \frac{d}{ds} \left[U_{T}(1,s) \ket{\tilde{ \phi}_0(s)}\right ]ds  
\end{align}
After differentiating by parts, we have
\begin{align}\label{delta1}
\delta(T)= \int_0^1 ds  \frac{d}{ds}\left[U_{T}(1,s)\right] \ket{\tilde{ \phi}_0(s)} + \int_0^1 ds U_{T}(1,s)\  \frac{d}{ds}  \ket{\tilde{ \phi}_0(s)} 
\end{align}
Inserting Schr\"{o}dinger Eq.~(\ref{SchrodingerUT}) into Eq.~(\ref{delta1})   we have
\begin{align}\nonumber
\delta(T)=&iT\int_0^1 ds   U_{T}(1,s) \hat{\tilde{H}}(s)  \ket{\tilde{ \phi}_0(s)} +T\int_0^1 ds U_{T}(1,s)\  \frac{d}{ds}  \ket{\tilde{ \phi}_0(s)} \\\label{delta18}
 =&T\int_0^1 ds U_{T}(1,s)\  \frac{d}{ds}  \ket{\tilde{ \phi}_0(s)}\;,
\end{align}
where we simplify the expression with the fact that $  \ket{\tilde{ \phi}_0(s)}  $ is the zero eigenvalue instantaneous eigenstate of the rotated Hamiltonian. To simplify the second term, we differentiate Eq.~(\ref{groundstate}) with respect to $ s $ on both side to obtain:
\begin{align}
&\frac{d \hat{\tilde{H}}(s) }{ds}\ket{\tilde{ \phi}_0(s)} + \hat{\tilde{H}}(s) \frac{d}{ds} \ket{\tilde{ \phi}_0(s)}=0,\\
\to &  \hat{\tilde{H}}(s) \frac{d}{ds} \ket{\tilde{ \phi}_0(s)}= -\frac{d  }{ds}\left[  \hat{H}_0 + U_{\dd}(0,s)\hat{\mathbb{H}}_{\od}(s))U_{\dd}(s,0)\right]\ket{\tilde{ \phi}_0(s)}=  -\frac{d  \hat{\tilde{H}}_{\od}(s) }{ds} \ket{\tilde{ \phi}_0(s)},\\\label{dsphi}
\to & \frac{d}{ds} \ket{\tilde{ \phi}_0(s)} = - \hat{G}^\prime \frac{d  \hat{\tilde{H}}_{\od}(s) }{ds}  \ket{\tilde{ \phi}_0(s)}
\end{align}
where we use the fact that $ \hat{H}_0 $ is a constant and the short hand notation $  \hat{\tilde{H}}_{\od}(s)=U_{\dd}(0,s)\hat{\mathbb{H}}_{\od}(s))U_{\dd}(s,0)$ and  the inverse of the effective Hamiltonian $ \hat{G}^\prime=\hat{\tilde{H}}^{-1} $ as
\begin{align}
 \hat{G}^\prime= \sum_{n\neq 0}\frac{1}{\tilde{E}_n}\ket{\phi_n(s)}\bra{\phi_n(s)}\approx   \sum_{n\neq 0}\frac{1}{E_n}\ket{\phi_n(s)}\bra{\phi_n(s)} + O(\frac{\epsilon^4}{\Delta^3})\approx \hat{G}\;,
\end{align}
where we use the same order of magnitude analysis as Eq.~(\ref{instantaneousBasis}): the rotated basis energy gap between different energy subspaces is the same as the original energy gap up to   the  order $O(\frac{\epsilon^4}{\Delta^3})  $ with $ \epsilon\ll \Delta $. Insert Eq.~(\ref{SchrodingerUT}) and Eq.~(\ref{dsphi}) into Eq.~(\ref{delta18}) and apply integration by parts, we obtain:
\begin{align}
\delta(T)=&\int_0^1\frac{1}{i T} \frac{d U_{T}(1,s) }{ds}\hat{\tilde{H}}^{-1}(-\hat{G}^\prime \frac{d  \hat{\tilde{H}}_{\od}(s) }{ds} ) \ket{\tilde{ \phi}_0(s)},\\\label{delta24}
=&\frac{i}{ T} U_{T}(1,s)(  \hat{G}^\prime)^2 \frac{d  \hat{\tilde{H}}_{\od}(s) }{ds}|_{s=0}^{s=1} - \frac{i}{T} \int_0^1 ds U_{T}(1,s) \frac{d}{ds}\left[ (  \hat{G}^\prime)^2 \frac{d  \hat{\tilde{H}}_{\od}(s) }{ds} \ket{\tilde{ \phi}_0(s)} \right].
\end{align}
Now using the results in Ref.~\cite{Jordan2008}, it is straight forward to give the bound on the magnitude of the adiabatic deviation given in Eq.~(\ref{delta24}) as
\begin{align}\nonumber
||\delta(T)|| \leq & \frac{1}{T}\left[ || ( \hat{G}^\prime)^2 \frac{d  \hat{\tilde{H}}_{\od}(s) }{ds}||_{s=0} +|| (\hat{G}^\prime)^2 \frac{d  \hat{\tilde{H}}_{\od}(s) }{ds}||_{s=1} +  \int_0^1 || \frac{d}{ds}  (  \hat{G}^\prime)^2 \frac{d  \hat{\tilde{H}}_{\od}(s) }{ds} \ket{\tilde{ \phi}_0(s)} ||ds
  \right]\\\nonumber
  &\leq \frac{1}{T} \left[  || ( \hat{G}^\prime)^2 \frac{d  \hat{\tilde{H}}_{\od}(s) }{ds}||_{s=0} +|| (\hat{G}^\prime)^2 \frac{d  \hat{\tilde{H}}_{\od}(s) }{ds}||_{s=1}  \right] \\\label{deltaTbound}
  &+   \frac{1}{T} \left[ \int_0^1 ds \left(   5|| (  \hat{G}^\prime)^3|| \, || \frac{d  \hat{\tilde{H}}_{\od}(s) }{ds}||^2 + || (  \hat{G}^\prime)^2|| \, || \frac{d^2  \hat{\tilde{H}}_{\od}(s) }{ds^2}||    \right) 
\right ].
\end{align}
Expanding the derivative of the off-diagonal Hamiltonian in the rotated basis we have
\begin{align}\label{dHoftilde}
\frac{d  \hat{\tilde{H}}_{\od}(s) }{ds}&=\frac{d   }{ds} \left( U_{\dd}(0,s)\hat{\mathbb{H}}_{\od}(s)U_{\dd}(s,0) \right),\\\nonumber
&= i T U_{\dd}(0,s)  \hat{\mathbb{H}}_{\dd}(s)\hat{\mathbb{H}}_{\od}(s)U_{\dd}(s,0) + U_{\dd}(0,s)\frac{d \hat{\mathbb{H}}_{\od}(s)}{ds} U_{\dd}(s,0) \\\nonumber
& - iTU_{\dd}(0,s)\hat{\mathbb{H}}_{\od}(s) \hat{\mathbb{H}}_{\dd}(s)U_{\dd}(s,0), \\\nonumber
&= i T U_{\dd}(0,s)[  \hat{\mathbb{H}}_{\dd}(s), \hat{\mathbb{H}}_{\od}(s)]U_{\dd}(s,0) + U_{\dd}(0,s)\frac{d \hat{\mathbb{H}}_{\od}(s)}{ds} U_{\dd}(s,0) .
\end{align}
This and the triangle inequality gives us an upper bound on the norm of the term:
\begin{align}\label{ofFirstDerivative}
|| \frac{d  \hat{\tilde{H}}_{\od}(s) }{ds}|| \leq  ||\frac{d \hat{\mathbb{H}}_{\od}(s)}{ds} || +T ||[  \hat{\mathbb{H}}_{\dd}(s), \hat{\mathbb{H}}_{\od}(s)]|| .
\end{align}
Differentiate  Eq.~(\ref{dHoftilde}) with respect to  $ s $  once more we obtain:
\begin{align}\nonumber
&\frac{d^2  \hat{\tilde{H}}_{\od}(s) }{ds^2}= 
-T^2 U_{\dd}(0,s)  \hat{\mathbb{H}}_{\dd}^2(s)\hat{\mathbb{H}}_{\od}(s)U_{\dd}(s,0) + i T U_{\dd}(0,s) \frac{d  \hat{\mathbb{H}}_{\dd}(s) }{ds} \hat{\mathbb{H}}_{\od}(s)U_{\dd}(s,0)  \\\nonumber
&+  i T U_{\dd}(0,s)  \hat{\mathbb{H}}_{\dd}(s) \frac{d \hat{\mathbb{H}}_{\od}(s)}{ds}U_{\dd}(s,0) +T^2U_{\dd}(0,s)   \hat{\mathbb{H}}_{\dd}(s)\hat{\mathbb{H}}_{\od}(s) \hat{\mathbb{H}}_{\dd}(s)U_{\dd}(s,0)\\\nonumber
 &+ i T U_{\dd}(0,s)  \hat{\mathbb{H}}_{\dd}(s)\frac{d \hat{\mathbb{H}}_{\od}(s)}{ds} U_{\dd}(s,0) +U_{\dd}(0,s)\frac{d^2 \hat{\mathbb{H}}_{\od}(s)}{ds^2} U_{\dd}(s,0)  - iT U_{\dd}(0,s)\frac{d \hat{\mathbb{H}}_{\od}(s)}{ds}  \hat{\mathbb{H}}_{\dd}(s) U_{\dd}(s,0)  \\\nonumber
 &+  T^2U_{\dd}(0,s) \hat{\mathbb{H}}_{\dd}(s)\hat{\mathbb{H}}_{\od}(s) \hat{\mathbb{H}}_{\dd}(s)U_{\dd}(s,0) - iTU_{\dd}(0,s)\frac{d \hat{\mathbb{H}}_{\od}(s)}{ds} \hat{\mathbb{H}}_{\dd}(s)U_{\dd}(s,0) \\\nonumber
 & - iTU_{\dd}(0,s)\hat{\mathbb{H}}_{\od}(s)\frac{d  \hat{\mathbb{H}}_{\dd}(s)}{ds} U_{\dd}(s,0)  -  T^2U_{\dd}(0,s)\hat{\mathbb{H}}_{\od}(s)\hat{H}^2_{\dd}(s)U_{\dd}(s,0)\\\nonumber
 &= T^2  U_{\dd}(0,s)  [ \hat{\mathbb{H}}_{\dd}(s), [ \hat{\mathbb{H}}_{\dd}(s), \hat{\mathbb{H}}_{\od}(s)]] U_{\dd}(s,0) + i 2T U_{\dd}(0,s) [ \hat{\mathbb{H}}_{\dd}(s),\frac{d \hat{\mathbb{H}}_{\od}(s)}{ds} ]U_{\dd}(s,0) \\
 &+U_{\dd}(0,s)\frac{d^2 \hat{\mathbb{H}}_{\od}(s)}{ds^2} U_{\dd}(s,0) 
 + i 2T U_{\dd}(0,s)[ \frac{d  \hat{\mathbb{H}}_{\dd}(s) }{ds} ,\hat{\mathbb{H}}_{\od}(s) ]U_{\dd}(s,0) 
\end{align}
This give us the upper bound on the second derivative in the non-adiabatic deviation as:
\begin{align}\label{offdiag2ndderivative}
|| \frac{d^2  \hat{\tilde{H}}_{\od}(s) }{ds^2}|| \leq & T^2 ||[ \hat{\mathbb{H}}_{\dd}(s), [ \hat{\mathbb{H}}_{\dd}(s), \hat{\mathbb{H}}_{\od}(s)]] || + 2T|| [ \hat{\mathbb{H}}_{\dd}(s),\frac{d \hat{\mathbb{H}}_{\od}(s)}{ds} ]|| \\\nonumber
&+ 2T ||[ \frac{d  \hat{\mathbb{H}}_{\dd} (s) }{ds} ,\hat{\mathbb{H}}_{\od}(s) ]|| +|| \frac{d^2 \hat{\mathbb{H}}_{\od}(s)}{ds^2} ||.
\end{align}
Inserting Eq.~(\ref{ofFirstDerivative}) and Eq.~(\ref{offdiag2ndderivative}) into Eq.~(\ref{deltaTbound}) completes the proof.

%

\vspace{20pt}

We now evaluate the magnitudes of each term of our non-adiabatic leakage bound in   off-resonant regime. Under the same assumption as our analysis for coherent leakage errors, we can  bound the order of magnitudes of non-adiabatic leakage errors from  Eq.~(\ref{nonadiabaticLeakageBound})  by $  \Vert\hat{\mathbb{H}}_{\dd}  \hat{\mathbb{H}}_{\od}  \Vert/  \Delta^2 \sim  O( \epsilon^4/\Delta^4 )$. It is  one order lower than that from the direct coupling leakage bound in Eq.~(\ref{totalLeakageBoundMaintext})  which is of the order $  \Vert \hat{\mathbb{H}}_{\od}  \Vert/ \Delta \sim O(\epsilon^3/\Delta^3)  $. As the result, off-resonant leakage errors are dominated by the direct coupling leakage errors. 
 
%

 \section{Control Filter Design}\label{app:cfd}
 
In this section, we describe in greater detail the design of the control filters used to suppress the first term of the  leakage error bounded  of  Eq.~(3) in the main text. Each time-dependent control  trajectory is discretized into  $ N $ steps in piece-wise constant representation, with neighboring steps separated by a small time duration $ \Delta t $. The modulation rate of the control signal is determined by this smallest time step according to $ F_{\text{sample}}=1/\Delta t $. We treat $ \hbar=1 $ throughout our discussion. Since the first term of leakage bound from Eq.~(3)
\begin{align}
\int_0^1 \frac{1}{\Delta^2(s)}\frac{1}{T}\left|\left| \frac{d^2 \hat{\mathbb{H}}_{\od}(s)}{ds^2} \right|\right|ds
\end{align}
is proportional to the second time-derivative of the block-off-diagonal Hamiltonian after TSWT, to suppress  it to a small value, we need to ensure that the  frequency component of  any control action proposed by RL agent proposed is sufficiently small compared to the energy gap $ \Delta $.  

Without affecting the Markovianity of the control problem while limiting the frequency components of the control, we choose to apply a  two pole normalized double exponential smoothing  filter~\cite{smith2007filter}  to the proposed control action. At each time step $1 \leq n\leq N $, the new control  depends not only on  the proposed control action $ \vec{c}^{RL}_n $ by the RL agent, but also   on  the control taken in the last step $ \vec{c}_{n-1} $ and the last last step $ \vec{c}_{n-2} $ according to:
\begin{align}
\vec{c}_n= a_1 \vec{c}^{RL}_n  - b_1\vec{c}_{n-1}- b_2\vec{c}_{n-2}\;,
\end{align}
 where the filter coefficient is chosen according to the   frequency bandwidth $ B_w $ of the actuated control and the modulation rate of the control signals as~\cite{smith2007filter}:
 \begin{align}
& \alpha=\exp[-\frac{\pi B_w}{ F_{\text{sample}}}],\\
 &a_1= (1-\alpha)^2, \, b_1=-2\alpha, \, b_2=\alpha^2.
 \end{align}
     
\section{Evaluation of the  Average Fidelity }\label{aveFidapp}
     
In this section, we present our method of evaluating  average fidelity based on previous results of Nielsen~\cite{nielsen2002simple}. The average fidelity,
 \begin{align}
 \bar{F}( \mathcal{E} , U)= \int d\psi\bra{\psi}U^\dagger \mathcal{E}(\ket{\psi}\bra{\psi})U\ket{\psi},
 \end{align}
measures the average performance  of a quantum gate over a uniform distribution of the input quantum state under the noisy quantum channel $ \mathcal{E} $ that describes the actual  control realizations.  It can be measured in experiment through randomized benchmarking process. Theoretically we   evaluate the average fidelity   by summing over all possible overlap with Pauli operators as
\begin{align}
&\bar{F}( \mathcal{E} , U)=\frac{\sum_j \text{Tr}[U_{\text{target}} U_j^\dagger U_{\text{target}}^\dagger \mathcal{E}(U_j)] +d^2}{d^2(d+1)},
\end{align}
where the two-qubit Pauli operator $ U_j=(\sigma_1^x)^{h_1}(\sigma_1^z)^{q_1}(\sigma_2^x)^{h_2}(\sigma_2^z)^{h_q} $ with $ h_j,q_j\in\{0,1\} $ satisfies $ \text{Tr}[U_j U_k^\dagger]= \delta_{j,k}  d$;  $ d=4 $ is the dimension of the two-qubit subspace dimension;  $  \mathcal{E}$ is the trace-preserving quantum operation that  represents   noisy realization of an ideal quantum gate control;  $ U_{\text{target}} $ represents the  target quantum gate.  It is not hard to see that the sufficient and necessary condition for average fidelity to be one is to have: $ \mathcal{E}(\rho)=U_{\text{target}}\rho U_{\text{target}}^\dagger $. We evaluate  $ \mathcal{E}(\rho) $ of a given control noise model $  N(0,\sigma_{\text{noise}}) $, a Gaussian random fluctuation in control amplitudes with zero mean and a fixed variance $ \sigma_{\text{noise}} $, by sampling the full control trajectory $ \vec{c}+ \delta \vec{c} $ under this noise channel and average the evolved quantum state  over  different instances:
\begin{align}
 \mathcal{E}(\rho)= \mathbb{E}_{ \delta \vec{c} \sim  N(0,\sigma_{\text{noise}})} U(\vec{c}+ \delta \vec{c})\rho  U^\dagger(\vec{c}+ \delta \vec{c}).
\end{align}
Such averaging can change a pure state  into a mixed state and thus accounts for the  magnitude of  decoherence  induced by  stochastic control errors. Fig.~4 of the main text is evaluated under $ 60 $ samples per noise model parameter.
 
 \end{appendix}

\end{document}